\documentclass[aps,prx,twocolumn,english,balance,superscriptaddress,floats,showpacs,letter, prb]{revtex4-2}
\usepackage[latin9]{inputenc}
\usepackage{amsmath}
\usepackage{amssymb}
\usepackage{graphicx}
\usepackage{esint}
\usepackage{subfigure}
\usepackage{multirow}
\usepackage{mathtools}
\usepackage{xcolor}
\usepackage{gensymb}

\makeatletter

\newcommand{\beq}{\begin{equation}}
\newcommand{\eeq}{\end{equation}}
\newcommand{\bea}{\begin{eqnarray}}
\newcommand{\eea}{\end{eqnarray}}
\newcommand{\bwt}{\begin{widetext}}
\newcommand{\ewt}{\end{widetext}}
\@ifundefined{textcolor}{}
{%
 \definecolor{BLACK}{gray}{0}
 \definecolor{WHITE}{gray}{1}
 \definecolor{RED}{rgb}{1,0,0}
 \definecolor{GREEN}{rgb}{0,1,0}
 \definecolor{BLUE}{rgb}{0,0,1}
 \definecolor{CYAN}{cmyk}{1,0,0,0}
 \definecolor{MAGENTA}{cmyk}{0,1,0,0}
 \definecolor{YELLOW}{cmyk}{0,0,1,0}
}

\newcommand{\eps}{\epsilon}
\newcommand{\veps}{\varepsilon}
\newcommand{\bk}{\mathbf{k}}
\newcommand{\bq}{\mathbf{q}}
\newcommand{\bp}{\mathbf{p}}
\newcommand{\bd}{\mathbf{d}}
\newcommand{\bg}{\mathbf{g}}
\newcommand{\br}{\mathbf{r}}

\newcommand{\bL}{\mathbf{L}}
\newcommand{\bK}{\mathbf{K}}
\newcommand{\bR}{\mathbf{R}}

\newcommand{\cT}{\mathcal{T}}

\newcommand{\fvec}[1]{\boldsymbol{#1}}

\newcommand{\rmd}{{\rm d}}

\makeatother

\begin{document}
	
\title{Lattice model for the Coulomb interacting chiral limit of the magic angle twisted bilayer graphene: symmetries, obstructions and excitations}

\author{Oskar Vafek}
\email{vafek@magnet.fsu.edu}
\affiliation{National High Magnetic Field Laboratory, Tallahassee, Florida, 32310, USA}
\affiliation{Department of Physics, Florida State University, Tallahassee, Florida 32306, USA}

\author{Jian Kang}
\email{jkang@suda.edu.cn}
\affiliation{School of Physical Science and Technology {\normalfont \&} Institute for Advanced Study, Soochow University, Suzhou, 215006, China}

\begin{abstract}
We revisit the localized Wannier state description of the twisted bilayer graphene, focusing on the chiral limit. We provide a simple method for constructing such 2D exponentially localized -- yet valley polarized -- Wannier states, centered on the sites of the honeycomb lattice, paying particular attention to maintaining all the unobstructed symmetries. This includes the unitary particle-hole symmetry, and the combination of $C_2\cT$ and the chiral particle-hole symmetry. The $C_2\cT$ symmetry alone remains topologically obstructed and is not represented in a simple site-to-site fashion.
We also analyze the gap and the dispersion of single particle and single hole excitations above a strong coupling ground state at integer fillings, which we find to be dominated by the on-site and the nearest neighbor terms of a triangular lattice hopping model, with a minimum at the center of the moire Brillouin zone. Finally, we use the insight gained from this real space description to understand the dependence of the gap and the effective mass on the range of the screened Coulomb interaction.
\end{abstract}

\maketitle
\section{Introduction}

There is a not uncommon sentiment that lattice models are a'priori unsuitable for the study of  the narrow bands of twisted bilayer graphene due to the topological nature of the narrow bands, and that extended state basis, such as Bloch or hybrid Wannier states, are preferable.
On the other hand, because the narrow bands~\cite{BMModel,Pablo1,Pablo2,Cory1,David,Young} as a whole within each valley carry a zero Chern number, the results of Brouder et.al.~\cite{MarzariPRL2007} guarantee that the exponentially localized Wannier states can be constructed, unlike if the Chern number were non-zero. The problem is very similar to the one studied by Soluyanov and Vanderbilt~\cite{SoluyanovPRB2011} where they explicitly constructed exponentially localized Wannier states for the  $Z_2$ non-trivial topological insulator whose Chern number of course vanishes. They also discovered that the non-trivial topological nature of the bands does provide an obstruction to the exponential localization of the Wannier states, {\it if} they also insist on simple on-site representation of the time reversal symmetry.  
For the example of the Kane and Mele model with two sites per unit cell and two pairs of bands separated by a non-trivial gap, the exponentially localized Wannier states for the two lower bands, say, can be constructed. But if we insist that the action of the time reversal on the Wannier states simply results in another Wannier state within the orthonormal set spanning the isolated Hilbert space, then the Wannier states cannot be exponentially localized.
If instead, the two Wannier states are centered on different sites of the unit cell and so do not transform into each other under the time reversal, then there is no obstruction to exponential localization.
It is often stated that the price to pay for the exponential localization is the breaking of the symmetry, which in the case of the 2D $Z_2$ topological insulator is the time reversal symmetry. But because the transformation between the Bloch states and the Wannier states is perfectly unitary, no information can be lost. This information includes symmetry.
Thus, strictly speaking, time reversal symmetry cannot be broken by such a unitary transformation, but rather it is not explicit (or on-site) in such Wannier representation. 
In practice this means that acting with a time reversal symmetry operator on an exponentially localized Wannier state will lead to a linear combination of Wannier states from its vicinity, with coefficients that decay exponentially with distance~\cite{WangPRB2020}. 

The situation is similar in the continuum models of magic angle twisted bilayer graphene whose narrow bands are taken to be completely valley decoupled, in which case they are topologically non-trivial~\cite{SenthilPRX2018,SongPRL2019,BABTBGI}.
In this case, exponentially localized states can be constructed~\cite{KangVafekPRX2018, LiangPRX2018} but if we insist on the explicit (on-site) implementation of both the valley $U(1)$ and $C_2\cT$ symmetry for the narrow bands, then we find an obstruction. 
This obstruction was originally thought to be fragile and removable by the addition of trivial remote bands~\cite{PoPRB2019}. But, as was recently shown in Ref.~\cite{BABTBGII}, including the particle-hole symmetry $P$ of the continuum Hamiltonian makes the topology stable, in that it is impossible to add trivial remote bands and maintain explicit representation of all approximate emergent symmetries while keeping exponential localization. 
In the chiral limit~\cite{Grisha} there is an additional chiral particle-hole symmetry $\mathcal{C}$ which will also play a role in our Wannier construction and the representation of a spin-valley $U(4)$ symmetry.

Although we agree that there is a clear utility of the Bloch~\cite{MacDonaldPRL2020, AshvinPRR2021, GuineaPNAS2018, SauPRB2020, NickPRX2020, DaiPRB2021, ZhangPRB2020} (or hybrid Wannier~\cite{KangVafekPRB2020, LeonPRB2021, ZaletelPRB2020}) description, one of the goals of this paper is to demonstrate the practical benefits of the Wannier states as a complementary description of the strong coupling physics of the narrow bands~\cite{KangVafekPRL2019, BrunoPRL2019}. In particular, we use them to explain the (exact) dispersion of a single particle or a single hole added to a correlated insulator at an integer filling in the strong coupling limit~\cite{VafekKangPRL2020,BABTBGV} of the chiral limit model~\cite{Grisha}. As was recently shown~\cite{VafekKangPRL2020}, the chiral limit is approached within the renormalization group treatment of the Coulomb interacting Bistritzer-MacDonald model~\cite{BMModel} of the magic angle twisted bilayer graphene.

Even though the $C_2\cT$ symmetry is not explicit in our Wannier basis, we construct the Wannier basis such that the combination of the chiral particle-hole symmetry $\mathcal{C}$ and $C_2\cT$ is explicit, as is $P$. The possibility to explicitly represent $P\mathcal{C}C_2\cT$ was insightfully pointed out by Bernevig et.al. in Ref.~\cite{BABTBGIII}. As we demonstrate below, the strong coupling spin-valley $U(4)$ symmetry, explicit in our Wannier basis, is indeed tied to $P\mathcal{C}C_2\cT$ as stated in Ref.~\cite{BABTBGIII}. 
To this end, we use the projection method~\cite{Vanderbilt} to construct the exponentially localized Wannier states by identifying suitable trial states, whose symmetry is then imprinted in the Wannier states. Usually one follows the projection method with a maximal localization procedure that minimizes $\langle \delta\br^2 \rangle$, but for our trial states it is not obvious that this second step provides sufficient enough advantage and therefore we omit it.

The strong coupling dispersion curves can certainly be obtained numerically without much difficulty using the Bloch basis, but it is not easy to understand their shape. Our goal is to demonstrate that the real space Wannier description provides a complementary way to understand it as a result of a tight-binding model on a moire {\it triangular lattice}, with rapidly decaying hopping amplitudes as the hopping range is increased~\cite{KangVafekCascades}. Note that the centers of the Wannier states are still on the honeycomb lattice sites, $AB$ and $BA$ as in Refs.~\cite{KangVafekPRX2018, LiangPRX2018}. The triangular lattice tight-binding description of the one-particle strong coupling dispersion is a result of decoupling of the two (triangular) sublattices of the honeycomb lattice in the chiral limit. 

The negative sign of the nearest neighbor hopping amplitude, which can also be understood from the real space description, then explains why the minimum of the dispersion is at the $\fvec\Gamma$ point (center of the moire Brillouin zone). In addition, this method allows us to understand the dependence of the shape of these bands on the range of the interaction, controlled by the distance $\xi$ to the grounded metallic screening layers, one placed above and another symmetrically below the twisted bilayer graphene. We find that once $\xi\gtrsim L_m$, where $L_m$ is the moire period, the effective mass of the excitations is nearly $\xi$-independent even though the excitation gap still grows with increasing $\xi$ (see Fig.~\ref{Fig:TBM}). This can be understood by analyzing the effective electrostatics of the exchange and the direct contribution to the single particle dispersion. The direct term corresponds to an electrostatic energy of a localized ``fidget spinner'', whose shape is shown in Fig.~\ref{Fig:WannierOverlaps:1}, interacting with a moire {\it periodic} charge distribution, thus probing the interaction potential $V_\bq$ only at (non-zero) discrete reciprocal lattice wavevectors. Because the smallest such wavevector has a length $|\bg_{1,2}|=4\pi/\left(\sqrt{3}L_m\right)\approx 7.26/L_m$, the factor of $\tanh\frac{|\bg_{1,2}|\xi}{2}$ in the two gate screened interaction $V_\bq=\frac{2\pi e^2}{\eps |\bq|}\tanh\left(\frac{|\bq|\xi}{2}\right)$ is not very sensitive to $\xi$ for $\xi\gtrsim L_m$.  This explains the $\xi$-insensitivity shown in Fig.~\ref{Fig:TBM:Hartree} for $\xi\gtrsim L_m$. 
In contrast, the exchange term corresponds to the electrostatic energy of {\it two localized} charge distributions, and thus probes the interaction at a continuum of wavevectors, not just at the discrete reciprocal lattice. By normalization, only the Wannier states centered on the same site --with the fidget spinner shape-- contain a monopole charge contribution which does not vanish at small wavevectors (see Fig.~\ref{Fig:WSMom:1}). The contribution of two fidget spinners to the exchange energy is therefore sensitive to small wavevector behavior of the interaction, which is the reason for the $\xi$-dependence of the on-site term in the tight binding expansion of the strong coupling dispersion seen in Fig.~\ref{Fig:Hopping:1}. On the other hand, the contribution to the hopping must contain at least one product of two Wannier states centered at different sites, whose multipole expansion generically starts with a dipole (see Fig.~\ref{Fig:WannierOverlaps:2}-\ref{Fig:WannierOverlaps:6}), vanishing at small wavevectors (see Fig.~\ref{Fig:WSMom:2}-\ref{Fig:WSMom:3}). The electrostatic energy is therefore much less sensitive to the small wavevector part of the interaction, and is therefore nearly independent of $\xi$ once $\xi\gtrsim L_m$.

Another goal of this paper is to elucidate the connection between the strong coupling $U(4)\times U(4)$ symmetry~\cite{NickPRX2020} in the chiral limit, the spin-valley $U(4)$ symmetry away from the chiral limit\cite{NickPRX2020} but assuming $P$ (dubbed non-chiral flat in Ref.~\cite{BABTBGIII}), and the local spin-valley $U(4)$ symmetry explicit in the Wannier basis~\cite{note}.
We demonstrate that last two $U(4)$ symmetries are indeed different subgroups of $U(4)\times U(4)$, as also pointed out in Ref.~\cite{BABTBGIII}.

This paper is organized as follows: 
in section II we define the starting model and spell out the action of its symmetries on the energy and Chern Bloch basis of the narrow bands. In section III we analyze the Coulomb interactions in the Bloch basis, display the (non-chiral flat) spin-valley $U(4)$ and (first chiral) $U(4)\times U(4)$ symmetries, and relate them. We also solve for the strong coupling spectrum of either a single particle or a single hole added to the strong coupling eigenstates at integer fillings using the Bloch basis. In section IV we build the exponentially localized Wannier states using the projection method and analyze the action of symmetries on the Wannier states; we also identify the spin-valley $U(4)$ symmetry of the projected density operator explicit in the Wannier basis and relate it to the (first chiral) $U(4)\times U(4)$ symmetry. In section V we analyze the strong coupling dispersion of either a single particle or a single hole added to the strong coupling eigenstates at integer fillings in the Wannier basis, and find the rapidly converging tight-binding hopping amplitudes for this dispersion at each integer filling $\nu$. As shown in Fig.~\ref{Fig:TBM:Fock}-\ref{Fig:TBM:Hartree} the results from the Bloch and Wannier analysis match. In section VI we analyze the dependence of the tight binding parameters on $\xi$ the range of the interaction and provide the electrostatics analogy to understand the direct and exchange contributions.
Finally, the section VII is devoted to summary and conclusions.

\section{Bistritzer-MacDonald Hamiltonian and its symmetries}

\begin{figure}[htbp]
	\centering
	\subfigure[\label{Fig:Schematic:Lattice}]{\includegraphics[width=0.54\columnwidth]{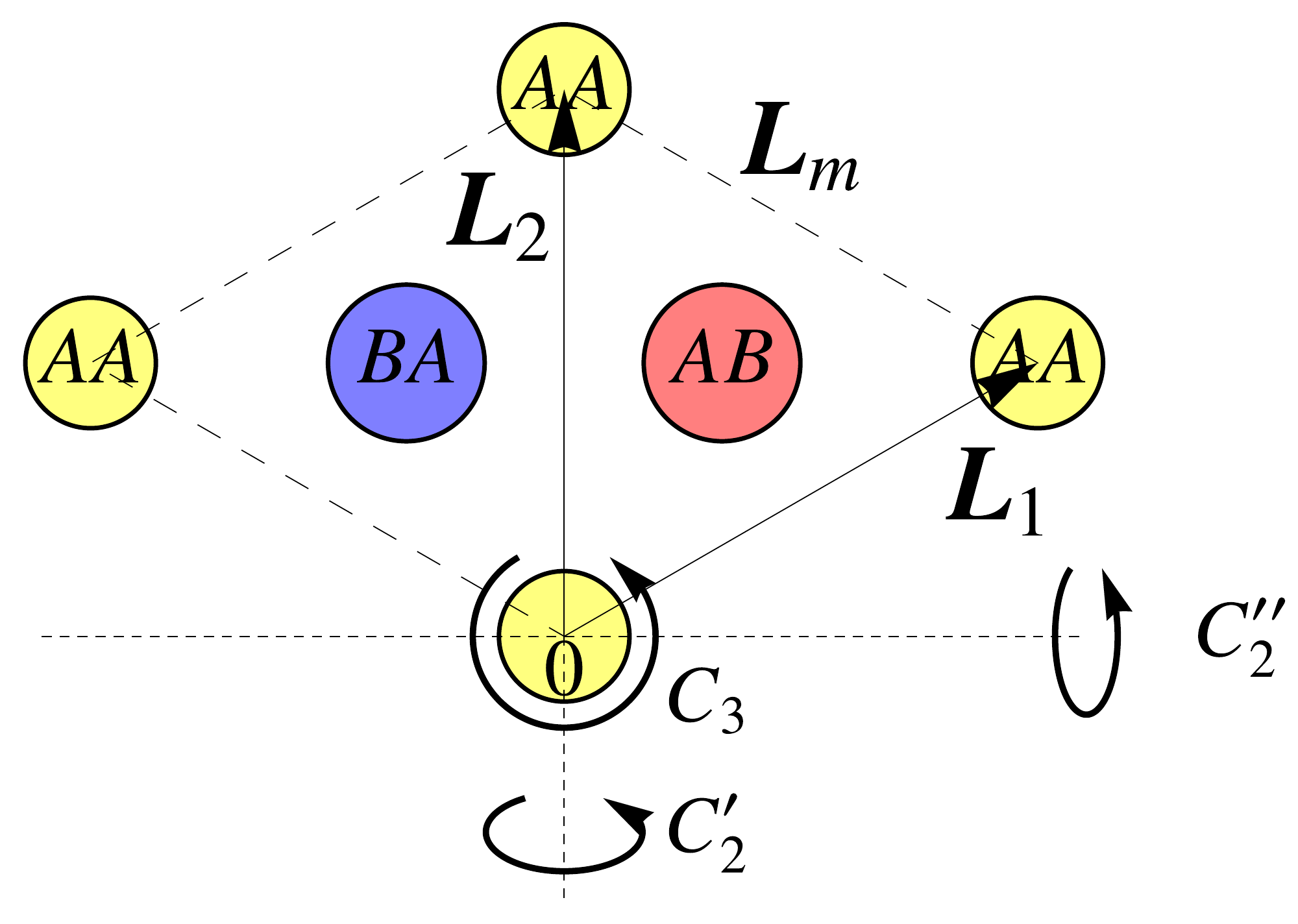}}
	\subfigure[\label{Fig:Schematic:BZ}]{\includegraphics[width=0.43\columnwidth]{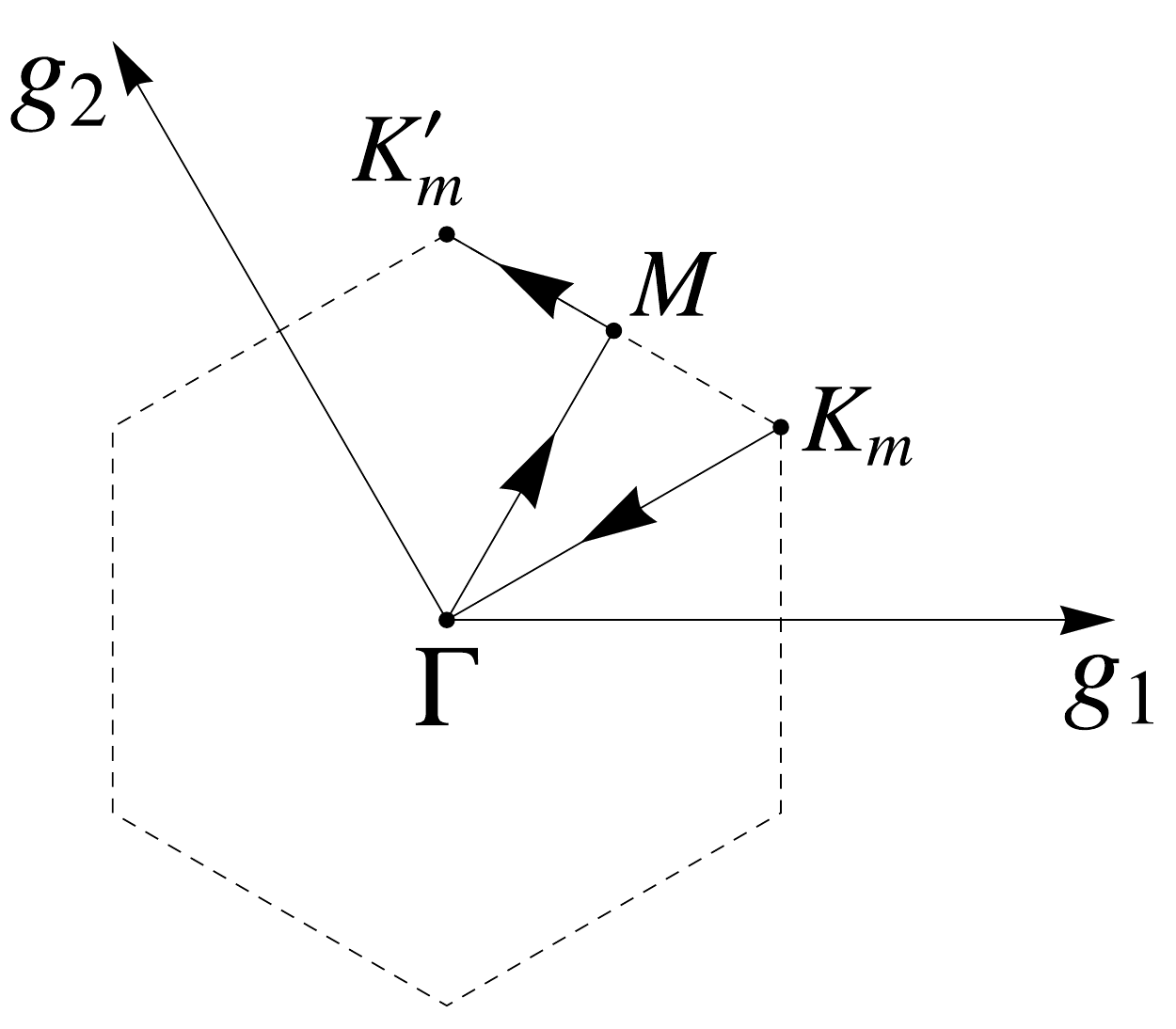}}
	\subfigure[\label{Fig:Schematic:Band}]{\includegraphics[width=0.95\columnwidth]{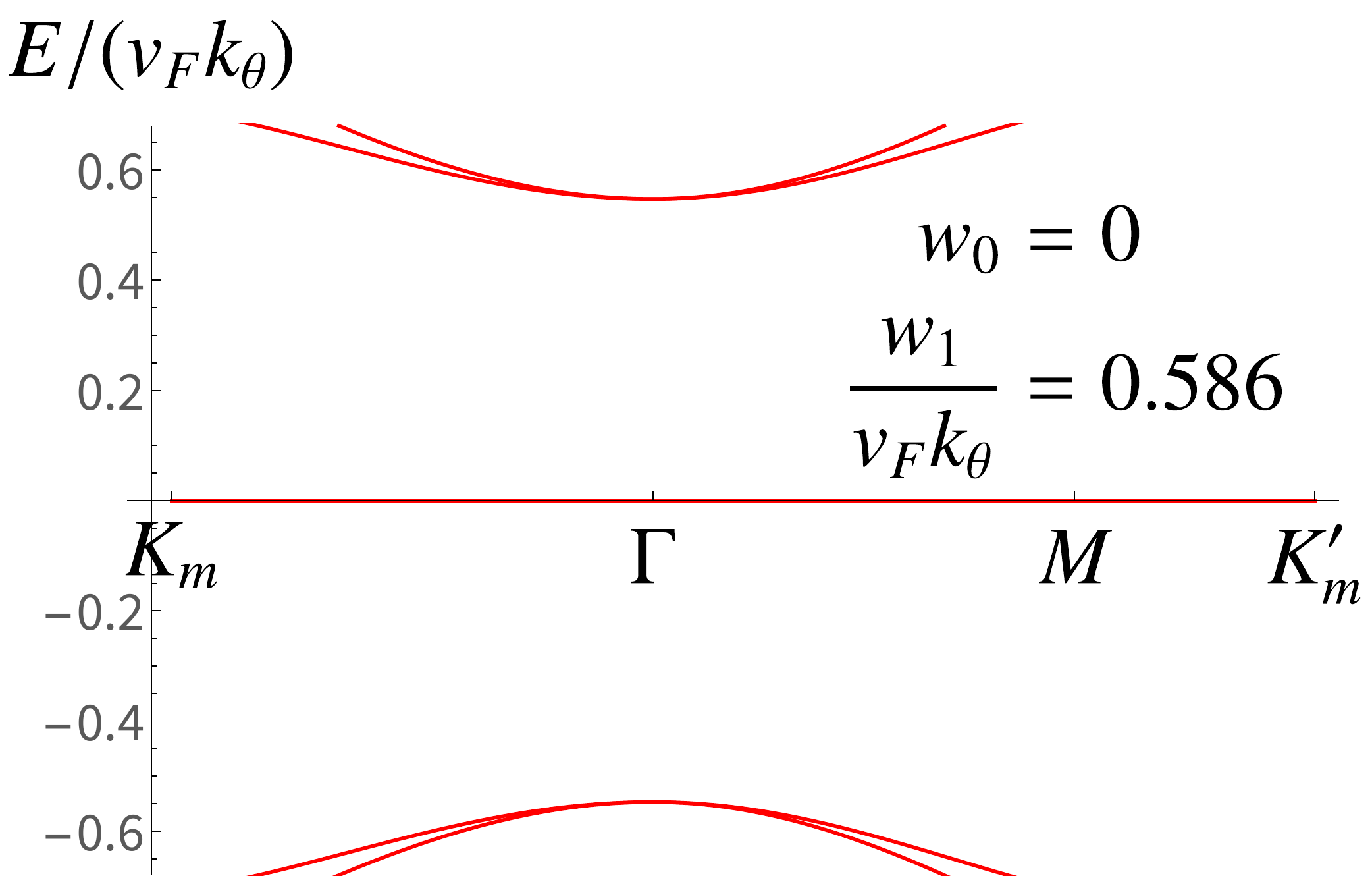}}
	\caption{(a) A unit cell of the triangular moire lattice with the schematic of the AA stacking regions. The constructed Wannier states are centered on the AB and BA stacking regions, forming the dual honeycomb lattice. (b) The high symmetry path of the moire Brillioun zone. (c) The narrow bands and a few remote bands in the chiral limit when $w_0 = 0$ and $w_1/(v_F k_{\theta})=0.586$. The non-interacting energy in panel (c) is calculated from BM model and expressed in the units of $v_F k_{\theta}$.}
	\label{Fig:Schematic}
\end{figure}

At the valley $\bK$ the effective continuum Hamiltonian is~\cite{BMModel, Grisha}
\begin{eqnarray}\label{eqn:HBM}
\hat{H}_{BM} &=&\left(\begin{array}{cc} v_F\sigma_\frac{\theta}{2}\cdot\bp & T(\br) \\
T^\dagger(\br) & v_F\sigma_{-\frac{\theta}{2}}\cdot\bp
\end{array}\right),
\end{eqnarray}
where the twisted Pauli matrices are $\sigma_\frac{\theta}{2}=e^{-\frac{i}{4}\theta\sigma_z}(\sigma_x,\sigma_y)e^{\frac{i}{4}\theta\sigma_z}$; they act in the sublattice space.
The interlayer hopping functions are
$T(\br)=\sum_{j=1}^3 T_j e^{-i\bq_j\cdot\br}$
where $\bq_1=k_\theta(0,-1)$, $\bq_{2,3}=k_\theta\left(\pm\frac{\sqrt{3}}{2},\frac{1}{2}\right)$, $k_\theta=\frac{8\pi}{3a_0}\sin\frac{\theta}{2}=4\pi/(3L_m)$, $a_0 \approx 0.246$nm, $L_m$ is the period of the moire lattice,
and
\begin{equation}
T_{j+1}=w_01_2+w_1\left(\cos\left(\frac{2\pi}{3}j\right)\sigma_x+\sin\left(\frac{2\pi}{3}j\right)\sigma_y\right),
\end{equation}
where $1_n$ is an $n\times n$ unit matrix. The Hamiltonian in the valley $\bK'$ is related to $\bK$ by spinless time reversal symmetry implemented by the complex conjugation.

We can absorb the phase in $\sigma_{\pm\frac{\theta}{2}}$ into the eigenstates using a unitary transformation~\cite{Grisha}
on the Hamiltonian
\begin{eqnarray}
\hat{H}'_{BM}=
\left(\begin{array}{cc}e^{\frac{i}{4}\theta\sigma_z} & 0 \\ 0 & e^{-\frac{i}{4}\theta\sigma_z} \end{array}\right)
\hat{H}_{BM}
\left(\begin{array}{cc}e^{-\frac{i}{4}\theta\sigma_z} & 0 \\ 0 & e^{\frac{i}{4}\theta\sigma_z} \end{array}\right).
\end{eqnarray}
In the chiral limit, $w_0/w_1=0$, this has no effect on $T(\br)$.
The eigenstates of the $\hat{H}'_{BM}$ can be written as
\begin{equation}\label{eqn:Psi}
\Psi_{n,\bk}(\br)=\sum_{\bg}\frac{1}{\sqrt{N_{uc}A_{uc}}}\left(\begin{array}{c} A_{n,\bg}(\bk)\\ B_{n,\bg}(\bk)e^{i\bq_1\cdot\br}\end{array}\right)e^{i\bk\cdot\br}e^{i\bg\cdot\br}
\end{equation}
where $\bg=m_1\bg_1+m_2\bg_2$ for integer $m_{1,2}$, $\bg_1=\bq_2-\bq_3$ and $\bg_{2}=\bq_{3}-\bq_1$ (see Fig.~\ref{Fig:Schematic}); $N_{uc}$ is the number of unit moire cells in the system (or equivalently, the number of $\bk$-points in the regular mesh of the first Brillouin zone), and the area of the unit cell is $A_{uc}=8\pi^2/(3\sqrt{3}k^2_\theta)$.
The Fourier coefficients are normalized as
$\sum_\bg A^\dagger_{n,\bg}(\bk)A_{n',\bg}(\bk)+B^\dagger_{n,\bg}(\bk)B_{n',\bg}(\bk)=\delta_{nn'}$.
The triangular moire lattice is spanned by the primitive lattice vectors $\bL_1=L_m(\frac{\sqrt{3}}{2},\frac{1}{2})$ and $\bL_2=L_m(0,1)$.

\subsection{Symmetries}
\label{sec:Symmetries Bloch}

\begin{figure}[htbp]
	\centering
	\subfigure[\label{Fig:TBM:Fock}]{\includegraphics[width=0.9\columnwidth]{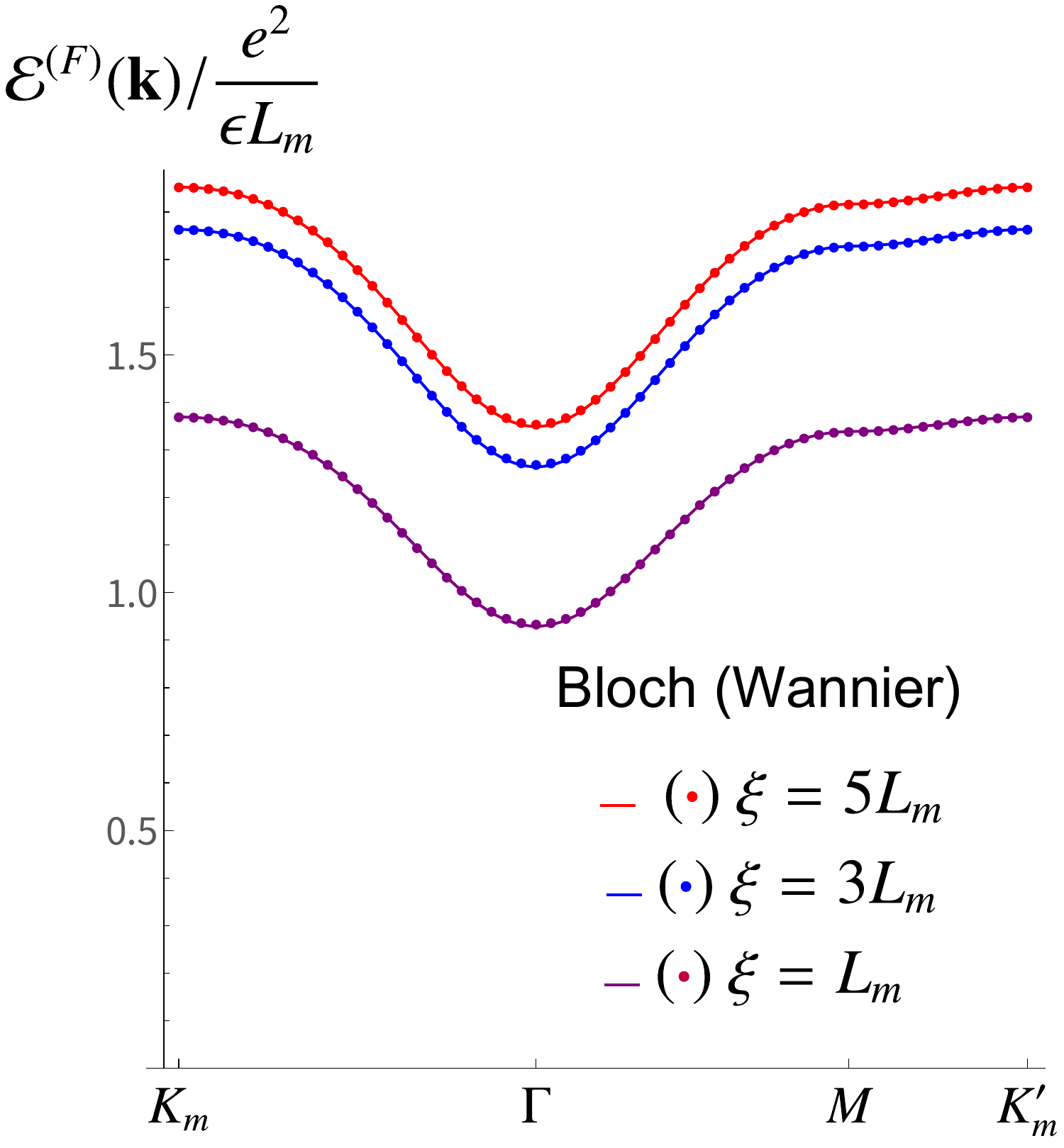}}
	\subfigure[\label{Fig:TBM:Hartree}]{\includegraphics[width=0.9\columnwidth]{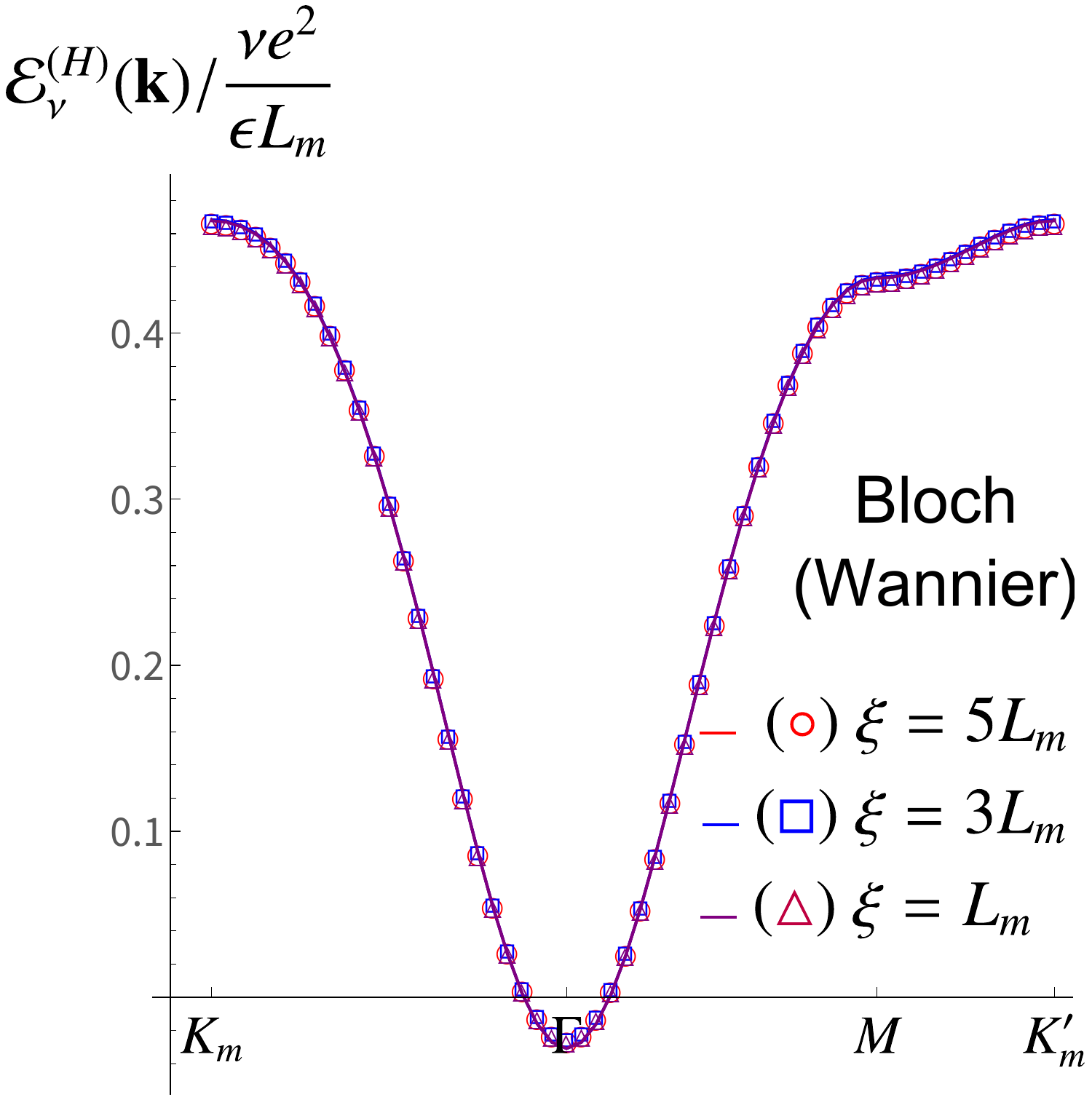}}
	\caption{The exchange (above) and the direct (below) contributions to the single particle dispersion in the strong coupling limit along the high symmetry path shown in Fig.~\ref{Fig:Schematic:BZ} for different values of $\xi$, the distance between the TBG and the metallic gates. $L_m$ is the period of the triangular moire lattice and the energy is expressed in Coulomb units $e^2/(\eps L_m)$. The solid colored curves are given by Eqn.~\ref{Eqn:BlochDisp} using Bloch states. The dots and other symbols denote the dispersion obtained from Wannier based tight binding model on triangular lattice with hopping distances no longer than $2L_m$ (Eqs.~\ref{eqn:HhpWan},\ref{Eqn:FockHopping} and \ref{Eqn:HartreeHopping}). Note that unlike the exchange contribution, the direct contribution in the lower pannel is almost independent of $\xi$ for $\xi \gtrsim L_m$.}
	\label{Fig:TBM}
\end{figure}

As emphasized by H.~C.~Po et.~al.~\cite{SenthilPRX2018}, the $\hat{H}_{BM}$ as well as $\hat{H}'_{BM}$ are invariant under the $C_2\cT$ symmetry i.e.
\begin{eqnarray}
C_2\cT:\;\;1_2\sigma_x{\hat{H}}'^{*}_{BM}(-\br)1_2\sigma_x &=& \hat{H}'_{BM}(\br),
\end{eqnarray}
where the two-by-two identity matrix $1_2$ in $1_2\sigma_x$ acts in the layer space.
This transformation does not change $\bk$ (or $n$), and we choose the phase of $\Psi_{n,\bk}(\br)$ such that it is an eigenstate of $C_2\cT$ with an eigenvalue $+1$, i.e.
\begin{equation}\label{eqn:C2TevalOne}
C_2T:\; 1_2\sigma_x\Psi^*_{n,\bk}(-\br)=\Psi_{n,\bk}(\br).
\end{equation}
This fixes the $\Psi_{n,\bk}(\br)$ up to an overall sign. For the purposes of this paper we will not need to fix the sign globally as we did in Ref.~\cite{KangVafekPRB2020}.

At a general twist angle away from the magic angle, the spectrum is non-degenerate everywhere in the mBZ except at the $\bK_m$ and $\bK'_m$ points even in the chiral limit~\cite{Grisha} defined as $w_0/w_1=0$. At a magic angle -- at the first of which $w_1/v_Fk_\theta= 0.58566355838955$, see Ref.~\cite{Becker2020}  -- and in the chiral limit, the narrow bands of $\hat{H}_{BM}$ are exactly flat~\cite{Grisha}. When we include the Coulomb interaction coupling to the remote bands within the renormalization group scheme, the perfect sublattice polarization of the narrow band Hilbert space remains, but the bands are no longer exactly flat~\cite{VafekKangPRL2020}.

For $w_0/w_1=0$, the $\hat{H}_{BM}$ and $\hat{H}'_{BM}$ enjoy the unitary chiral particle-hole symmetry~\cite{Grisha}
\begin{eqnarray}
\mathcal{C}:\;\;\left\{1_2\sigma_z,\hat{H}'_{BM}\right\}=0.
\end{eqnarray}
Therefore, if $\Psi_{n+,\bk}(\br)$ is an eigenstate of $\hat{H}'_{BM}$ with an eigenvalue $\eps_{n+,
\bk}$, then $1_2\sigma_z\Psi_{n+,\bk}(\br)$ is an eigenstate with an eigenvalue $-\eps_{n+,\bk}=\eps_{n-,\bk}$; the two eigenvalues are distinct unless we are exactly at a magic angle or $\bk=\bK_m,\bK'_m$.
This means that the linear combinations
\begin{eqnarray}\label{eqn:Phi}
\Phi_{\pm,\bk}(\br)&=&\frac{1}{\sqrt{2}}\left(1_2\sigma_z \pm 1_4\right)\Psi_{n+,\bk}(\br)
\end{eqnarray}
are perfectly sublattice polarized with $1_2\sigma_z$ eigenvalues $\pm1$, and thus carry the Chern indices of $\pm 1$ respectively~\cite{Grisha}. 
Note that Eq.\ref{eqn:Phi} is well defined for both signs because $C_2\cT$ anticommutes with $\mathcal{C}$, and we have already assumed (\ref{eqn:C2TevalOne}).
We also have
\begin{equation}\label{eqn:chiralPH}
\mathcal{C}:\; 1_2\sigma_z \Psi_{n+,\bk}(\br)=e^{i\alpha^\mathcal{C}_\bk}\Psi_{n-,\bk}(\br),
\end{equation}
where $\alpha^\mathcal{C}_\bk=\pm\frac{\pi}{2}$, which can be obtained by acting on both sides with $C_2\cT$
and using Eqn.\ref{eqn:C2TevalOne}.

Because in the strong coupling limit, and near the magic angle, the Coulomb interaction exceeds the narrow bandwidth (containing two bands for each valley and spin) it will generally mix the two bands.
Therefore, it is a matter of convenience whether we choose the Chern basis $\Phi_{\pm,\bk}(\br)$ which diagonalizes $1_2\sigma_z$ for the narrow bands, or $\Psi_{n\pm,\bk}(\br)$ which diagonalizes $\hat{H}'_{BM}$.

As emphasized by Z.~Song et.~al.~\cite{SongPRL2019} and K.~Hejazi et.~al.~\cite{BalentsPRB2019}, $\hat{H}'_{BM}$ also enjoys another unitary particle-hole symmetry (see also Refs.~\cite{NickPRX2020} and \cite{CanoPRR2021})
\begin{eqnarray}\label{eqn:P}
P:\; \left(\begin{array}{cc} 0 & -1_2 \\
1_2 & 0\end{array}\right){\hat{H}}'_{BM}(-\br)\left(\begin{array}{cc} 0 & 1_2 \\
-1_2 & 0\end{array}\right) &=& -\hat{H}'_{BM}(\br).\nonumber\\
\end{eqnarray}
Therefore, if $\Psi_{n,\bk}(\br)$ is an eigenstate of $\hat{H}'_{BM}$ with an eigenvalue $\eps_{n,\bk}$, then $-i\mu_y1_2 \Psi_{n,\bk}(-\br)$ is an eigenstate at $-\bk-\bq_1$ with the eigenvalue $-\eps_{n,\bk}$; the Pauli matrix $\mu_y$ acts in the layer space.
This symmetry is approximately present even away from the chiral limit when the twist angle is small~\cite{BABTBGI}, and becomes exact if we ignore the rotation of the Pauli matrices in Eqn.\ref{eqn:HBM}.
Because the particle-hole symmetry $P$ commutes with $C_2\cT$ and because we already fixed the $C_2\cT$ eigenvalue to $+1$ in Eqn.~\ref{eqn:C2TevalOne}, we have
\begin{equation}\label{eqn:PHpsi}
P:\; -i\mu_y1_2 \Psi_{n+,\bk}(-\br)=e^{i\alpha^P_\bk}\Psi_{n-,-\bk-\bq_1}(\br),
\end{equation}
where $\alpha^P_\bk=0,\pi$. We will find it convenient to fix the gauge such that
\begin{equation}
\alpha^P_\bk=0,\;\;\forall \bk\in\text{mBZ}.
\end{equation}
Then, at the $\fvec{M}$ and $\fvec{\Gamma}$ points there are two independent ways to arrive at $\Psi_{n-,\bk}$ starting from $\Psi_{n+,\bk}$, either using Eqn.~\ref{eqn:PHpsi} or using Eqn.~\ref{eqn:chiralPH}. At the $\fvec{M}$ point ($\bk=-\frac{\bq_1}{2}$), numerically we find that these are consistent if $\alpha^\mathcal{C}_{\fvec{M}}=-\frac{\pi}{2}$; we find the same result at the equivalent $\fvec{M}$ points $\bk=\frac{\bq_{2,3}}{2}$.
However, at the $\fvec{\Gamma}$ point ($\bk=\bq_1$), we find $\alpha^\mathcal{C}_{\fvec{\Gamma}}=\frac{\pi}{2}$.

We therefore use the remaining freedom to fix
\begin{eqnarray}
\alpha^\mathcal{C}_{\bk}&=&-\frac{\pi}{2},\;\;\bk\neq \fvec{\Gamma}\\
\alpha^\mathcal{C}_{\bk}&=&\frac{\pi}{2},\;\; \bk=\fvec{\Gamma}.
\end{eqnarray}
This locks the sign of $\Psi_{n+,-\bk-\bq_1}(\br)$ to $\Psi_{n+,\bk}(\br)$ according to
\begin{eqnarray}
\Psi_{n+,-\bk-\bq_1}(\br)&=&-\mu_y\sigma_z \Psi_{n+,\bk}(-\br),\;\; \bk\neq \fvec{\Gamma}\\
\Psi_{n+,-\bk-\bq_1}(\br)&=&\mu_y\sigma_z \Psi_{n+,\bk}(-\br),\;\; \bk= \fvec{\Gamma}.
\end{eqnarray}
Therefore,
\begin{eqnarray}\label{eqn:CforPhi}
\mathcal{C}&:&\; 1_2\sigma_z\Phi_{\pm,\bk}(\br)= \pm \Phi_{\pm,\bk}(\br),\\
\label{eqn:C2TforPhi}
C_2\cT &:&\; 1_2\sigma_x\Phi^*_{\pm,\bk}(-\br)=- \Phi_{\mp,\bk}(\br),\\
P &:&\; -i\mu_y1_2\Phi_{\pm,\bk}(-\br)=\pm e^{i\alpha^P_\bk}e^{-i\alpha^\mathcal{C}_{-\bk-\bq_1}}\Phi_{\pm,-\bk-\bq_1}(\br)\nonumber\\
&&=\pm e^{-i\alpha^\mathcal{C}_{-\bk-\bq_1}}\Phi_{\pm,-\bk-\bq_1}(\br).
\label{eqn:PforPhi}
\end{eqnarray}

In addition, $\hat{H}'_{BM}$ is also invariant under the $C'_2\cT$ and $C_3$ symmetries~\cite{SenthilPRX2018,KangVafekPRX2018,LiangPRX2018}
\begin{eqnarray}
C'_2\cT &:&\; \mu_x{\hat{H}}'^{*}_{BM}(-x,y)\mu_x = \hat{H}'_{BM}(x,y),\\
C_3&:&\; e^{-i\frac{\pi}{3}1_2\sigma_z}e^{-i\frac{2\pi}{3}\hat{\ell}_z}{\hat{H}}_{BM}e^{i\frac{2\pi}{3}\hat{\ell}_z}e^{i\frac{\pi}{3}1_2\sigma_z}=
{\hat{H}}_{BM},
\end{eqnarray}
where $\hat{\ell}_z=-i\left(x\frac{\partial}{\partial y}-y\frac{\partial}{\partial x}\right)$.
Because the spectrum is non-degenerate everywhere except at the $\bK_m$ and $\bK'_m$ points, the action of these symmetries maps the Bloch states at the related $\bk$ points up to an overall phase.

The two valleys are related by (spinless) time reversal i.e. by complex conjugation.
The remaining symmetries can then be generated by the combination of the ones stated above.

\section{Coulomb interactions}

The electron-electron Coulomb interactions, projected onto the narrow bands~\cite{VafekKangPRL2020, BABTBGIII}, are
\begin{eqnarray}\label{eqn:Hint}
&&H_{int}=\frac{1}{2}\int d^2\br d^2\br' V(\br-\br')\delta \rho(\br)\delta \rho(\br')=
\nonumber\\
&&\frac{1}{2N_{uc}A_{uc}}
\sum_\bq V_{\bq}\sum_{\bR\bR'}e^{i\bq\cdot(\bR-\bR')}
\delta O^\dagger_\bR(\bq)\delta O_{\bR'}(\bq),
\end{eqnarray}
where
\begin{eqnarray}\label{eqn:deltarho}
\delta \rho(\br)&=&\chi^\dagger_\sigma(\br)\chi_\sigma(\br)-2\sum_{\bk}\sum_{n=n\pm}\Psi^\dagger_{n,\bk}(\br)\Psi_{n,\bk}(\br),\nonumber\\
V(\br)&=&\frac{1}{N_{uc}A_{uc}}\sum_\bq V_{\bq}e^{i\bq\cdot\br},\\
\delta O_{\bR}(\bq) &=&
\int_{Wig.-Seitz} d^2\br
e^{-i\bq\cdot\br}
\delta \rho(\bR+\br).
\end{eqnarray}
In the above, we are keeping only $V_\bq$ with $|\bq|\ll |\bK|$, thus neglecting the large wavevector transfer terms in the Coulomb interaction which are smaller by a factor $\sim L_m|\bK|$~\cite{ZaletelPRL2020}.


For a pure Coulomb interaction we would have $V_{\bq}=2\pi e^2/(\eps q)$ except at $\bq=\fvec{0}$ where $V_{\bq=\fvec{0}}=0$; in the later sections we also study the two gate screened interactions, in which case $V_{\bq\neq \fvec{0}}=\frac{2\pi e^2}{\eps q}\tanh\frac{q\xi}{2}$.
The projected field operators in the Eq.(\ref{eqn:deltarho}) are expanded in the narrow band basis fermion annihilation
operators $d_{\sigma,\bK/\bK',n,\bk}$ as
\begin{eqnarray}\label{eqn:chiPsi}
\chi_\sigma(\br)&=&\left(\begin{array}{c} \psi_\sigma(\br) \\ \phi_\sigma(\br)\end{array}\right)
=\sum_{\bk}\sum_{n=n\pm}
\left(\begin{array}{c} \Psi_{n,\bk}(\br) d_{\sigma,\bK,n,\bk} \\ \Psi^*_{n,\bk}(\br) d_{\sigma,\bK',n,-\bk-\bq_1}
\end{array}\right).\nonumber\\
\end{eqnarray}

\begin{widetext}
If we now explicitly write out the projected density operator, using $C_2\cT$ (Eqn.~\ref{eqn:C2TevalOne}), $P$ (Eqn.~\ref{eqn:PHpsi}) and $\alpha^P_{\bk}=0$, we find that the terms from valley $\bK$ and $\bK'$ are related according to
\begin{eqnarray}\label{eqn:densityPC2T}
\chi^\dagger_\sigma(\br)\chi_\sigma(\br)&=&\sum_{\bk\bk'}
\left(d^\dagger_{\sigma,\bK,+,\bk},d^\dagger_{\sigma,\bK,-,\bk},d^\dagger_{\sigma,\bK',+,\bk},d^\dagger_{\sigma,\bK',-,\bk}\right)
\left(\begin{array}{cccc} A_{\bk\bk'}(\br) & B_{\bk\bk'}(\br) & 0 & 0 \\
C_{\bk\bk'}(\br) & D_{\bk\bk'}(\br) & 0 & 0\\
0 & 0 & D_{\bk\bk'}(\br) & -C_{\bk\bk'}(\br) \\
0 & 0 & -B_{\bk\bk'}(\br) & A_{\bk\bk'}(\br)
\end{array}\right)
\left(\begin{array}{c} d_{\sigma,\bK,+,\bk'} \\ d_{\sigma,\bK,-,\bk'} \\ d_{\sigma,\bK',+,\bk'} \\ d_{\sigma,\bK',-,\bk'}
\end{array}\right)\nonumber\\
\end{eqnarray}
where $A_{\bk\bk'}(\br)=\Psi^\dagger_{n+,\bk}(\br)\Psi_{n+,\bk'}(\br)$, $B_{\bk\bk'}(\br)=\Psi^\dagger_{n+,\bk}(\br)\Psi_{n-,\bk'}(\br)$, $C_{\bk\bk'}(\br)=\Psi^\dagger_{n-,\bk}(\br)\Psi_{n+,\bk'}(\br)=B^*_{\bk'\bk}(\br)$ and $D_{\bk\bk'}(\br)=\Psi^\dagger_{n-,\bk}(\br)\Psi_{n-,\bk'}(\br)$.
The above matrix commutes with $1_4$, $\tau_z1_2$, $\tau_y\tilde{\sigma}_y$ and $\tau_x\tilde{\sigma}_y$ where the Pauli matrices $\tau$ act in the valley space and $\tilde{\sigma}$ in the $n\pm$ band space~\cite{VafekKangPRL2020}. Together with the spin SU(2) generators, they form the 16 generators of the global spin-valley U(4) symmetry first pointed out by Bultinck et.al. Ref.~\cite{NickPRX2020}. At this point we have not used the $\mathcal{C}$ symmetry, only $C_2\cT$ and $P$, therefore, this U(4) symmetry is present even away from the chiral limit, if we neglect the small angle rotation of the Pauli matrices in $\hat{H}_{BM}$ (Eqn.\ref{eqn:HBM})~\cite{SongPRL2019}. We will adopt the nomenclature of Ref.~\cite{BABTBGIII} and refer to this as the non-chiral flat spin-valley $U(4)$ symmetry.

The consequences of the $\mathcal{C}$ symmetry are more transparent in the $\Phi$-basis (Eqn.~\ref{eqn:Phi}).
Expanding the fields as
\begin{eqnarray}\label{eqn:chiPhi}
\chi_\sigma(\br)&=&\left(\begin{array}{c} \psi_\sigma(\br) \\ \phi_\sigma(\br)\end{array}\right)
=\sum_{\bk}\sum_{\lambda=\pm}
\left(\begin{array}{c} \Phi_{\lambda,\bk}(\br) e_{\sigma,\bK,\lambda,\bk} \\ e^{i\alpha^{\mathcal{C}}_\bk}\Phi^*_{\lambda,\bk}(\br) e_{\sigma,\bK',-\lambda,-\bk-\bq_1}
\end{array}\right),
\end{eqnarray}
and using Eqns.(\ref{eqn:CforPhi})-(\ref{eqn:PforPhi}) we find
\begin{eqnarray}
\chi^\dagger_\sigma(\br)\chi_\sigma(\br)&=&\sum_{\bk\bk'}
\left(e^\dagger_{\sigma,\bK,+,\bk},e^\dagger_{\sigma,\bK,-,\bk},e^\dagger_{\sigma,\bK',-,\bk},e^\dagger_{\sigma,\bK',+,\bk}\right)
\left(\begin{array}{cccc} \bar{A}_{\bk\bk'}(\br) & 0 & 0 & 0 \\
0 & \bar{D}_{\bk\bk'}(\br) & 0 & 0\\
0 & 0 & \bar{D}_{\bk\bk'}(\br) & 0 \\
0 & 0 & 0 & \bar{A}_{\bk\bk'}(\br)
\end{array}\right)
\left(\begin{array}{c} e_{\sigma,\bK,+,\bk'} \\ e_{\sigma,\bK,-,\bk'} \\ e_{\sigma,\bK',-,\bk'} \\ e_{\sigma,\bK',+,\bk'}
\end{array}\right)\nonumber\\
\end{eqnarray}
where $\bar{A}_{\bk\bk'}(\br)=\Phi^\dagger_{+,\bk}(\br)\Phi_{+,\bk'}(\br)$ and $\bar{D}_{\bk\bk'}(\br)=\Phi^\dagger_{-,\bk}(\br)\Phi_{-,\bk'}(\br)$.
Clearly, in this basis, we can perform a global spin-valley $U(4)$ rotation in the Chern basis $\left(e_{\sigma,\bK,+,\bk},e_{\sigma,\bK',+,\bk}\right)$ and an independent spin-valley $U(4)$ rotation in the opposite Chern basis $\left(e_{\sigma,\bK,-,\bk},e_{\sigma,\bK',-,\bk}\right)$, leading to the spin valley $U(4)\times U(4)$ symmetry in the chiral limit first pointed out by Bultinck et.al.~\cite{NickPRX2020} (see also Ref.~\cite{VafekKangPRL2020, BABTBGIII}.)
Adopting the nomenclature of Ref.~\cite{BABTBGIII}, we will refer to this as the (first) chiral-flat $U(4)\times U(4)$ symmetry.
\end{widetext}

In the Appendix~\ref{apx:U4s}, we relate the generators of the non-chiral flat $U(4)$ to the generators of the (first) chiral flat $U(4)\times U(4)$ (as was also done in Ref.~\cite{NickPRX2020, BABTBGIII}). 
Performing an arbitrary non-chiral flat $U(4)$ global rotation is equivalent to rotating the Chern basis $\left(e_{\sigma,\bK,+,\bk},e_{\sigma,\bK',+,\bk}\right)$ by $\exp\left[i\left(\omega_0\tau_0+\omega_z\tau_z+\omega_1\tau_x+\omega_2\tau_y\right)\right]$
while simultaneously rotating the opposite Chern basis $\left(e_{\sigma,\bK,-,\bk},e_{\sigma,\bK',-,\bk}\right)$ by $\exp\left[i\left(\omega_0\tau_0+\omega_z\tau_z-\omega_1\tau_x-\omega_2\tau_y\right)\right]$.
This can be thought of a mirror reflecting an axial vector $\vec{\omega}$  about the xy-plane.

Note that if $\delta \rho(\br)$ acts on the $\nu=0$ fully valley polarized state, such as $|\Omega_{\nu=0}\rangle=\prod_{\bk,n=\pm}d^\dagger_{\uparrow,\bK,n,\bk}d^\dagger_{\downarrow,\bK,n,\bk}|0\rangle$,
then $\delta \rho(\br)$ annihilates it. Because $H_{int}$ is positive semi definite, this is therefore a ground state of $H_{int}$. Due to the $U(4)\times U(4)$ symmetry, any state obtained by such global rotation is also annihilated and is therefore degenerate with the valley polarized state.
Using $P$ (Eqn.\ref{eqn:PforPhi}), $C_2\cT$ (Eqn.\ref{eqn:C2TforPhi}) and Eqn.\ref{eqn:Phi}, we have
\begin{eqnarray}
\sum_{\bk}\sum_{n=\pm}\Psi^\dagger_{n,\bk}(\br)\Psi_{n,\bk}(\br)&=&2\sum_{\bk}\Phi^\dagger_{+,\bk}(\br)\Phi_{+,\bk}(\br)\\
&=&2\sum_{\bk}\Phi^\dagger_{-,\bk}(\br)\Phi_{-,\bk}(\br).
\end{eqnarray}
Therefore, for any integer filled Chern state $|\Omega_\nu\rangle$, we have
\begin{eqnarray}\label{eqn:rhoeval}
&&\delta \rho(\br)|\Omega_\nu\rangle = \delta \bar{\rho}_\nu(\br)|\Omega_\nu\rangle\\
&&\delta \bar{\rho}_\nu(\br)=\nu\sum_{\bk}\Phi^\dagger_{+,\bk}(\br)\Phi_{+,\bk}(\br)=\nu\sum_{\bk}\Phi^\dagger_{-,\bk}(\br)\Phi_{-,\bk}(\br)\nonumber\\
&&=\frac{\nu}{2}\sum_{\bk}\sum_{n=\pm}\Psi^\dagger_{n,\bk}(\br)\Psi_{n,\bk}(\br).
\label{eqn:rhobar}
\end{eqnarray}
The last equality will be useful when we switch to the 2D exponentially localized Wannier basis because it corresponds to the sum over all states of the two narrow bands whose total Chern number vanishes, for which the localized basis can always be found~\cite{MarzariPRL2007}.

\subsection{Excitations in the strong coupling}

In the strong coupling, the energy of an operator $X$ acting on the Chern state $|\Omega_\nu\rangle$ at integer filling $\nu$ can be obtained as
\begin{eqnarray}
&&E X|\Omega_\nu\rangle = H_{int}X|\Omega_\nu\rangle.
\label{eqn:eigeneqn}
\end{eqnarray}
Substituting (\ref{eqn:Hint}) and using (\ref{eqn:rhoeval}) we readily find
\begin{eqnarray}
&&\left(E-E^{(0)}_{\nu}\right) X|\Omega_\nu\rangle  \nonumber\\
&=&\frac{1}{2}\int d^2\br d^2\br' V(\br-\br')\left[\delta \rho(\br),\left[\delta \rho(\br'),X\right]\right]|\Omega_\nu\rangle\nonumber\\
&+&\int d^2\br d^2\br' V(\br-\br')\left[\delta \rho(\br),X\right]\delta \bar{\rho}_\nu(\br')|\Omega_\nu\rangle,
\label{eqn:disp com}
\end{eqnarray}
where $E^{(0)}_{\nu}=\frac{1}{2}\int d^2\br d^2\br' V(\br-\br')\delta \bar{\rho}_\nu(\br)\delta \bar{\rho}_\nu(\br')$ is a real number, and $\delta\bar\rho_\nu(\br)$ was defined in Eq.~\ref{eqn:rhobar}.
The key observation is that the commutators on the right hand side of Eq.~\ref{eqn:disp com} do not change the number of creation and annihilation operators in $X$ because $\delta\rho(\br)$, defined in Eq.~\ref{eqn:deltarho}, is a charge neutral bilinear. Therefore, we can just match the operator coefficients on each side of the equation to find their spectra (assuming again that $X$ does not annihilate $|\Omega_\nu\rangle$, in which case it is trivial).

\subsection{Single particle excitation using Bloch basis}

To facilitate the calculation, we note that if we rearrange the components as
\begin{equation}\mathfrak{d}_{\bk}=\left(\begin{array}{c} d_{\sigma,\bK,+,\bk} \\ d_{\sigma,\bK,-,\bk} \\ d_{\sigma,\bK',-,\bk}\\ -d_{\sigma,\bK',+,\bk}
\end{array}\right)
\end{equation}
then we can write the projected density in Eqn.~\ref{eqn:densityPC2T} as
\begin{eqnarray}
\chi_\sigma^\dagger(\br)\chi_\sigma(\br) =\sum_{\bk\bk'}
\mathfrak{d}^\dagger_{n,\bk,\sigma,\kappa}
\Psi^\dagger_{n,\bk}(\br)\Psi_{n',\bk'}(\br)
\mathfrak{d}_{n',\bk',\sigma,\kappa}\nonumber\\
\end{eqnarray}
where repeated subscripts are summed. The above expression is explicitly $U(4)$ symmetric.

To compute the energy spectrum of a single hole added to the strong coupling ground state we let $X=\mathfrak{d}_{n,\bk,\sigma,\kappa}$. For a single electron we let $X=\mathfrak{d}^\dagger_{n,\bk,\sigma,\kappa}$.
Such spectra were presented in Ref.~\cite{VafekKangPRL2020, BABTBGV}.
In addition, spectra for the charge neutral (excitonic) excitation were computed in Ref.~\cite{VafekKangPRL2020, Eslam2020,BABTBGV} and for the charge $\pm2$ excitations in Ref.~\cite{BABTBGV}.
We note in passing that in order to determine the degeneracy, we also need to consider the nature of the ground state to make sure $X$ does not annihilate it.
\begin{widetext}
Substituting into Eqn.~\ref{eqn:disp com}, and using the discrete translational invariance of the moire lattice, we find
\begin{eqnarray}
\left(E-E^{(0)}_{\nu}\right) \mathfrak{d}_{n,\bk,\sigma,\kappa}|\Omega_\nu\rangle &=&
\frac{1}{2}\int d^2\br d^2\br' V(\br-\br')
\sum_{m}\sum_{m'\bp'}  \Psi^\dagger_{n,\bk}(\br')\Psi_{m',\bp'}(\br')
\Psi^\dagger_{m',\bp'}(\br)\Psi_{m,\bk}(\br)
\mathfrak{d}_{m,\bk,\sigma,\kappa}
|\Omega_\nu\rangle\nonumber\\
&-&\frac{\nu}{2}\int d^2\br d^2\br' V(\br-\br')\sum_{m' \bp'}\sum_{m}  \Psi^\dagger_{m',\bp'}(\br')\Psi_{m',\bp'}(\br')
\Psi^\dagger_{n,\bk}(\br)\Psi_{m,\bk}(\br)
\mathfrak{d}_{m,\bk,\sigma,\kappa}|\Omega_\nu\rangle,\\
\left(E-E^{(0)}_{\nu}\right) \mathfrak{d}^\dagger_{n,\bk,\sigma,\kappa}|\Omega_\nu\rangle &=&
\frac{1}{2}\int d^2\br d^2\br' V(\br-\br')
\sum_{m}\sum_{m'\bp'}  \Psi^\dagger_{m',\bp'}(\br')\Psi_{n,\bk}(\br')
\Psi^\dagger_{m,\bk}(\br)\Psi_{m',\bp'}(\br)
\mathfrak{d}^\dagger_{m,\bk,\sigma,\kappa}
|\Omega_\nu\rangle\nonumber\\
&+&\frac{\nu}{2}\int d^2\br d^2\br' V(\br-\br')\sum_{m' \bp'}\sum_{m}  \Psi^\dagger_{m',\bp'}(\br')\Psi_{m',\bp'}(\br')
\Psi^\dagger_{m,\bk}(\br)\Psi_{n,\bk}(\br)
\mathfrak{d}^\dagger_{m,\bk,\sigma,\kappa}|\Omega_\nu\rangle.
\end{eqnarray}
We see that the right hand side is a linear combination of operators $\mathfrak{d}_{n,\bk,\sigma,\kappa}$ on the left hand side, which means that the above equation can be satisfied if we perform an appropriate unitary transformation on the operators on both sides.
Because the momentum $\bk$ of the operators on both sides is the same, as are the spin $\sigma$ and valley $\kappa$, to find the energy spectrum of the single particle excitations, we just need to diagonalize a $2\times 2$ matrix in the band indices $n,m$. The matrix is Hermitian due to the interaction potential $V(\br-\br')$ being even under particle exchange, which, together with $C_2\cT$ (Eqn.\ref{eqn:C2TevalOne}) allows us to show that this matrix differs for the hole and the particle only by the direct (Hartree) term
\begin{eqnarray}\label{eqn:Hhp}
\mathcal{H}^{h/p}_{nm}(\bk)&=&\frac{1}{2}\int d^2\br d^2\br' V(\br-\br')
\sum_{m'\bp'}  \Psi^\dagger_{n,\bk}(\br)\Psi_{m',\bp'}(\br)
\Psi^\dagger_{m',\bp'}(\br')\Psi_{m,\bk}(\br')\nonumber\\
&\mp&
\frac{\nu}{2}\int d^2\br d^2\br' V(\br-\br')\sum_{m' \bp'}\Psi^\dagger_{m',\bp'}(\br)\Psi_{m',\bp'}(\br)
\Psi^\dagger_{n,\bk}(\br')\Psi_{m,\bk}(\br').  \label{Eqn:BlochDisp}
\end{eqnarray}
\end{widetext}
The matrix $\mathcal{H}^{h/p}_{nm}(\bk)$ is diagonal in $nm$ for each $\bk$. This is because according to Eqn.~\ref{eqn:C2TevalOne} each $\Psi_{n,\bk}(\br)$ is a $C_2\cT$ eigenstate with an eigenvalue $+1$, and because we can use $\mathcal{C}$ to flip between the signs of $n$ utilizing Eqn.~\ref{eqn:chiralPH}. Moreover, $C_2\cT=1_2\sigma_x \mathcal{K}$ while $\mathcal{C}=1_2\sigma_z$. Therefore, $\mathcal{C}C_2\cT = i 1_2\sigma_y \mathcal{K}$ squares to $-1$. It is clearly anti-unitary and does not change $\bk$.
Therefore, at any $\bk$, $\mathcal{C}C_2\cT$ guarantees that
\begin{eqnarray}
\left(\begin{array}{cc}0 & 1 \\ -1 &0 \end{array}\right)_{nn'}{\mathcal{H}^*}^{h/p}_{n'm'}(\bk)\left(\begin{array}{cc}0 & -1 \\ 1 &0 \end{array}\right)_{m'm}&=& \mathcal{H}^{h/p}_{nm}(\bk).
\end{eqnarray}
Because it is a Hermitian $2\times 2$ matrix, the above guarantees that $\mathcal{H}^{h/p}_{nm}(\bk)$ is proportional to the identity matrix $1_2$ in the chiral limit, i.e. the two eigenvalues are degenerate at each $\bk$. The contribution from the first (exchange) and the second (direct, divided by $\mp\nu$) lines in Eqn.~\ref{Eqn:BlochDisp} are shown in the Fig.~\ref{Fig:TBM}. To obtain the actual degeneracies of the bands, we need to consider whether acting with a particular linear combination of $d$'s or $d^\dagger$'s annihilates the ground state (which, up to the $U(4)\times U(4)$ transformation is a product state of the Chern states). For example, for $\nu=1$ there are 3 ways to add a (light) particle and 5 ways to add a (heavy) hole~\cite{KangVafekCascades}. 
For a general non-negative integer filling $\nu\geq 0$, the particle branch is $(4-\nu)$-fold degenerate (because there are $4-\nu$ ways to add a particle without annihilating the many body product state) and the hole branch is $(4+\nu)$-fold degenerate (because there are $4+\nu$ ways to add a hole). The degeneracies for negative integer filling $\nu< 0$ are related by the particle-hole transformation: 
the hole branch is $(4-|\nu|)$-fold degenerate and the particle branch is $(4+|\nu|)$-fold degenerate.

Away from the chiral limit, $\mathcal{C}$ is no longer exact. However, to a good approximation~\cite{BABTBGII}, even away from the chiral limit we still have $P$. Because $PC_2\cT$ is also antiunitary and squares to $-1$, and because it changes $\bk$ to $-\bk-\bq_1$, $\mathcal{H}^{h/p}_{nm}(\bk)$ must be diagonal at all momenta left invariant under $P$, namely $\fvec\Gamma$ and $\fvec{M}$. Away from these $\bk$-points, the spectrum is split as shown in Refs.~\cite{VafekKangPRL2020,BABTBGV,KangVafekCascades}.
For a range of angles near the magic, evaluating the spectrum in the Bloch basis requires numerical determination of the wavefunctions $\Psi_{n,\bk}(\br)$,
although exactly at the magic angle and the chiral limit, one should only need the numerical determination of the wavefunctions at $\bK_m$\cite{Grisha}.
Nevertheless, the reason for the shape of the spectrum thus obtained is somewhat obscured in the Bloch basis.
To better reveal its character, we now switch to the Wannier basis.

\section{Wannier states in the chiral limit}
We now turn to the main part of the paper where we use the projection method~\cite{Vanderbilt} to construct the exponentially localized Wannier states in the chiral limit.
There are two Wannier states per unit cell in the valley $\bK$, namely $w_{\bR,j}(\br)$ with $j=1\equiv AB$ and $j=2\equiv BA$. For a fixed triangular moire lattice vector $\bR=m_1\bL_1+m_2\bL_2$ with integer $m_{1,2}$, they are distinguished by the position of their center on the dual honeycomb lattice which is either $AB$ or $BA$ (see Fig.~\ref{Fig:Schematic:Lattice}). The exponentially localized Wannier states can be obtained from the Fourier transform of an appropriate linear combination of the Bloch states $\Phi_{\pm,\bk}$ distinguished by their sublattice polarization (\ref{eqn:Phi}) as
\begin{eqnarray}\label{eqn:w def}
w_{\bR,j}(\br)&=& \frac{1}{\sqrt{N_{uc}}}\sum_{\bk} e^{-i\bk\cdot\bR}\tilde\Phi_{j,\bk}(\br),\\
\tilde\Phi_{j,\bk}(\br) &=&\sum_{m=\pm}\Phi_{m,\bk}(\br)\mathcal{U}_{mj}(\bk),\\
\mathcal{U}_{mj}(\bk)&=&\text{UP}\left[\int d^2\br \Phi^\dagger_{m,\bk}(\br)h_{j}(\br)\right].
\end{eqnarray}
In the above, UP stands for ``unitary part'' and $h_{j}(\br)$ are the trial functions (at each valley there are two trial functions; the trial functions at valley $\bK'$ are related to the trial functions at $\bK$ by spinless time reversal symmetry i.e. complex conjugation).
The most practical method for constructing the UP of the $\bk$-dependent $2\times 2$ matrix
\begin{equation}\label{eqn:overlapmatrixA}
A_{mj}(\bk)=\int d^2\br \Phi^\dagger_{m,\bk}(\br)h_{j}(\br)
\end{equation}
is via singular value decomposition (SVD), and replacing the diagonal part of the SVD with an identity matrix~\cite{Vanderbilt}. This procedure is justified provided none of the singular values of $A_{mj}(\bk)$ vanish at some $\bk$, otherwise there would be an obstruction to exponential localization of Wannier states~\cite{Vanderbilt}.

\begin{figure}[htbp]
	\centering
    \includegraphics[width=0.8\columnwidth]{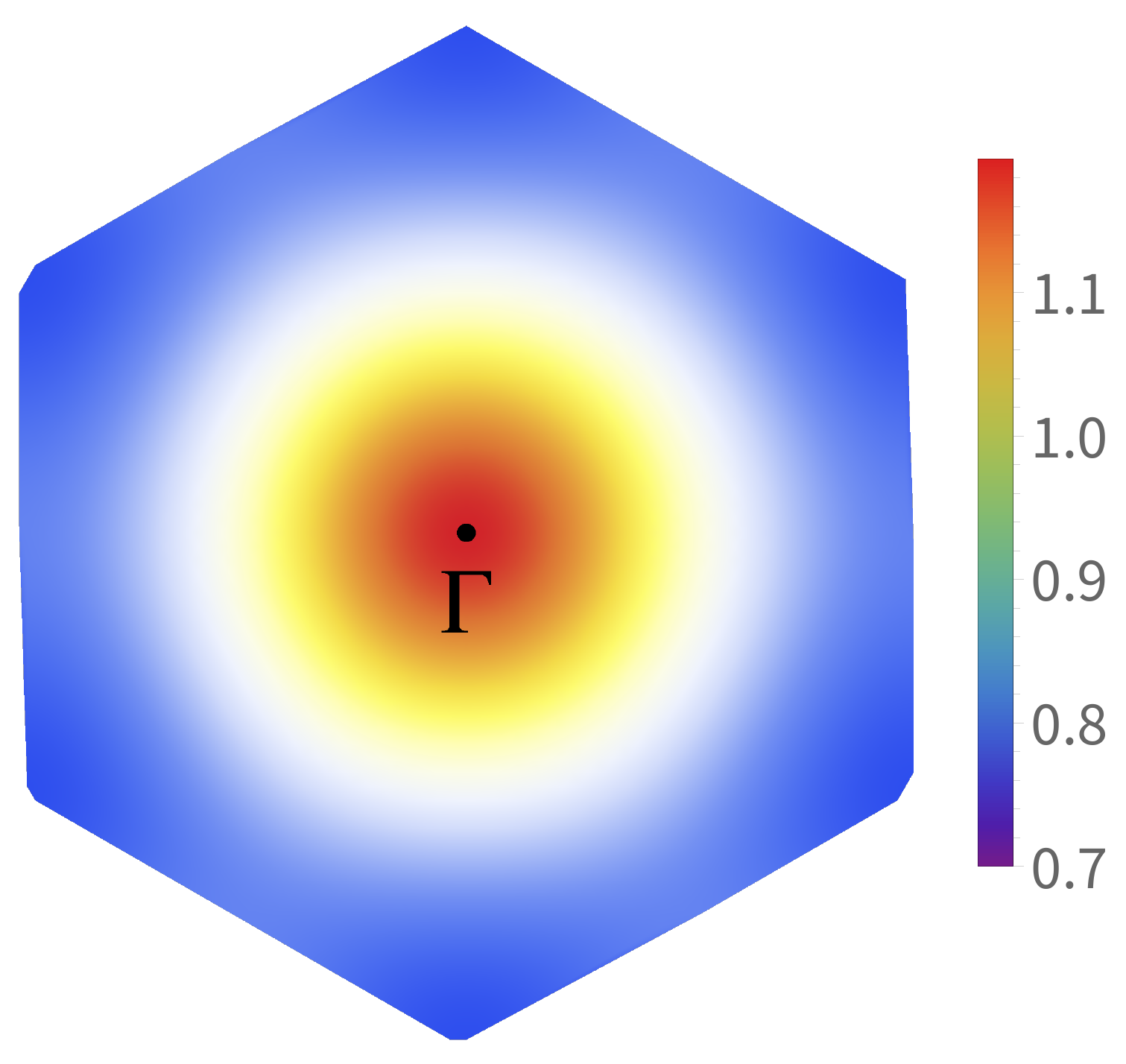}
	\caption{The (degenerate) singular values of the matrix $A_{mj}(\bk)$ defined in Eqn.~\ref{eqn:overlapmatrixA} with  the trial functions in Eqn.~\ref{eqn:h1} and \ref{eqn:h2}.}
	\label{Fig:SVD}
\end{figure}

A particularly appealing choice for the trial functions in the chiral limit $w_0/w_1=0$ is
\begin{eqnarray}\label{eqn:h1}
h_1(\br)&=&\left(\begin{array}{c} i \\ 0 \\ 0 \\ -\veps^*
\end{array}\right)\delta\left(\br-\br_{AB}\right),\\
\label{eqn:h2}
h_2(\br)&=&\veps^* C'_2\cT h_1(\br)=\left(\begin{array}{c} 0 \\ -1 \\ -i\veps^* \\ 0
\end{array}\right)\delta\left(\br-\br_{BA}\right),
\end{eqnarray}
where $\br_{AB}=\frac{1}{3}\left(\bL_1+\bL_2\right)$, $\br_{BA}=\frac{1}{3}\left(2\bL_2-\bL_1\right)$, $\veps=e^{2\pi i/3}=e^{-i\bq_1\cdot\br_{AB}}=e^{-i\bq_1\cdot\br_{BA}}$, and $\delta(\br)$ is the 2D Dirac $\delta$-function.
With this choice, the singular values of $A_{mj}(\bk)$ are degenerate, never vanish, and are relatively weakly $\bk$-dependent (see Fig.~\ref{Fig:SVD}).
This immediately implies that the Wannier states are 2D exponentially localized.

Moreover, not only do we guarantee the on-site representation of $C'_2\cT$ --which is incorporated by construction-- but, remarkably, we simultaneously obtain simple representation of $P$, $\mathcal{C}C_2\cT$ and $C_3$.  We demonstrate this in the Appendix~\ref{sec:ApxWan}.
The $C_2\cT$ alone remains obstructed and cannot be represented on-site.

\subsection{Action of the symmetries on the Wannier states}
 As detailed in the Appendix~\ref{sec:ApxWan}, the action of the particle-hole symmetry (\ref{eqn:P}) on the Wannier states (\ref{eqn:w def}) is
\begin{eqnarray}\label{eqn:P on w1}
P:\;\;&& -i\mu_y 1_2 w_{\bR,1}(-\br)=-\veps^* e^{i\bq_1\cdot\bR}w_{-\bR-\bL_2,2}(\br)\\
\label{eqn:P on w2}
&&-i\mu_y 1_2 w_{\bR,2}(-\br)
=\veps^*e^{i\bq_1\cdot\bR}w_{-\bR-\bL_2,1}(\br).
\end{eqnarray}
Similarly, we derive the action of $C_2\cT$ followed by $\mathcal{C}$ in the Appendix~\ref{sec:ApxWan} to be
\begin{eqnarray}\label{eqn:CC2T on w1}
\mathcal{C}C_2\cT: \;\; &&
i1_2\sigma_y  w^*_{\bR,1}(-\br)=-i w_{-\bR-\bL_2,2}(\br)\\
\label{eqn:CC2T on w2}
&&i1_2\sigma_y  w^*_{\bR,2}(-\br)
=i w_{-\bR-\bL_2,1}(\br).
\end{eqnarray}
Time reversal followed by the in-plane 2-fold rotation along $y$-axis gives
\begin{eqnarray}
C'_2\cT:\;\;&& \mu_x 1_2 w^*_{\bR,1}(-x,y)
=\veps  w_{-R_x,R_y,2}(x,y)\\
&&\mu_x 1_2 w^*_{\bR,2}(-x,y)=\veps  w_{-R_x,R_y,1}(x,y)
\end{eqnarray}
which follows using the similar arguments detailed in the Appendix and the definition of $h_{2}$ via the action of $C_2\cT$ on $h_1$ in Eqn.\ref{eqn:h2}.
Finally, the action of 3-fold rotations about the axis perpendicular to the plane on the Wannier states give
\begin{eqnarray}
C_3: \;\; &&
e^{-i\frac{\pi}{3}1_2\sigma_z} e^{-i\frac{2\pi}{3}\hat{\ell}_z} w_{\bR,1}(\br)=\nonumber\\
&&e^{-i\frac{\pi}{3}}w_{\left(-\frac{1}{2}R_x-\frac{\sqrt{3}}{2}R_y,-\frac{1}{2}R_y+\frac{\sqrt{3}}{2}R_x\right)-\bL_1,1}(\br)\\
&&e^{-i\frac{\pi}{3}1_2\sigma_z} e^{-i\frac{2\pi}{3}\hat{\ell}_z} w_{\bR,2}(\br)=\nonumber\\
&&e^{i\frac{\pi}{3}}w_{\left(-\frac{1}{2}R_x-\frac{\sqrt{3}}{2}R_y,-\frac{1}{2}R_y+\frac{\sqrt{3}}{2}R_x\right)-\bL_2,2}(\br).
\end{eqnarray}

Combining $P$ and $\mathcal{C} C_2\cT$ we have an unobstructed symmetry~\cite{BABTBGII} and
\begin{eqnarray}
 && w_{\bR,j}(\br)= i\mu_y \sigma_y \veps^*e^{i\bq_1\cdot\bR} w^*_{\bR,j}(\br).
\end{eqnarray}
Therefore, if we let
\begin{eqnarray}\label{eqn:wtilde}
\tilde{w}_{\bR,j}(\br)= e^{i\frac{\pi}{12}}e^{-\frac{i}{2}\bq_1\cdot\bR} w_{\bR,j}(\br)
\end{eqnarray}
then
\begin{eqnarray}
\tilde{w}_{\bR,j}(\br)= \mu_y \sigma_y \tilde{w}^*_{\bR,j}(\br)
\end{eqnarray}
and, we therefore reach an important conclusion, namely
\begin{equation}\label{eqn:WS overlap is real}
\tilde{w}^\dagger_{\bR,j}(\br)\tilde{w}_{\bR',j'}(\br)\in \text{Reals}.
\end{equation}
This is because expanding the fields in $\tilde{w}$ basis
\begin{eqnarray}
\chi_\sigma(\br)&=&
\sum_{\bR}\sum_{j=1,2}
\left(\begin{array}{c} \tilde{w}_{\bR,j}(\br) f_{\sigma,\bK,j,\bR} \\ \tilde{w}^*_{\bR,j}(\br) f_{\sigma,\bK',j,\bR}
\end{array}\right),
\end{eqnarray}
then gives the projected density operator
\begin{eqnarray}
&&\chi^\dagger_\sigma(\br)\chi_\sigma(\br)=\sum_{\bR,\bR'}\sum_{j,j'}\tilde{w}^\dagger_{\bR,j}(\br)\tilde{w}_{\bR',j'}(\br)\times
\nonumber\\
&&
\left(f^\dagger_{\sigma,\bK,j,\bR}f_{\sigma,\bK,j',\bR'}+f^\dagger_{\sigma,\bK',j,\bR}f_{\sigma,\bK',j',\bR'}\right),
\end{eqnarray}
which is explicitly invariant under a spin-valley $U(4)$ rotation. Here, the fermion operator $f_{\sigma,\kappa,j,\bR}$ annihilates the Wannier state at site $\bR$ sublattice $j$, spin $\sigma$ and valley $\kappa=\bK,\bK'$.
In the Appendix \ref{apx:U4s} we relate this $U(4)$ subgroup, dubbed chiral non-flat $U(4)$ in Ref.~\cite{BABTBGIII}, to the (first) chiral-flat limit $U(4)\times U(4)$ symmetry. 
Performing an arbitrary chiral non-flat $U(4)$ global rotation is equivalent to simultaneously rotating the aforementioned Chern bases $\left(e_{\sigma,\bK,+,\bk},e_{\sigma,\bK',+,\bk}\right)$ and $\left(e_{\sigma,\bK,-,\bk},e_{\sigma,\bK',-,\bk}\right)$ by the {\it same} transformation, say $\exp\left[i\left(\omega_0\tau_0+\omega_z\tau_z+\omega_1\tau_x+\omega_2\tau_y\right)\right]$. This is unlike in the non-chiral flat $U(4)$ where the axial vector $\vec{\omega}$ was reflected about the xy-plane.

\begin{widetext}
\begin{figure*}[t]
	\centering
	\subfigure[\label{Fig:WannierOverlaps:1}]{\includegraphics[width=0.6\columnwidth]{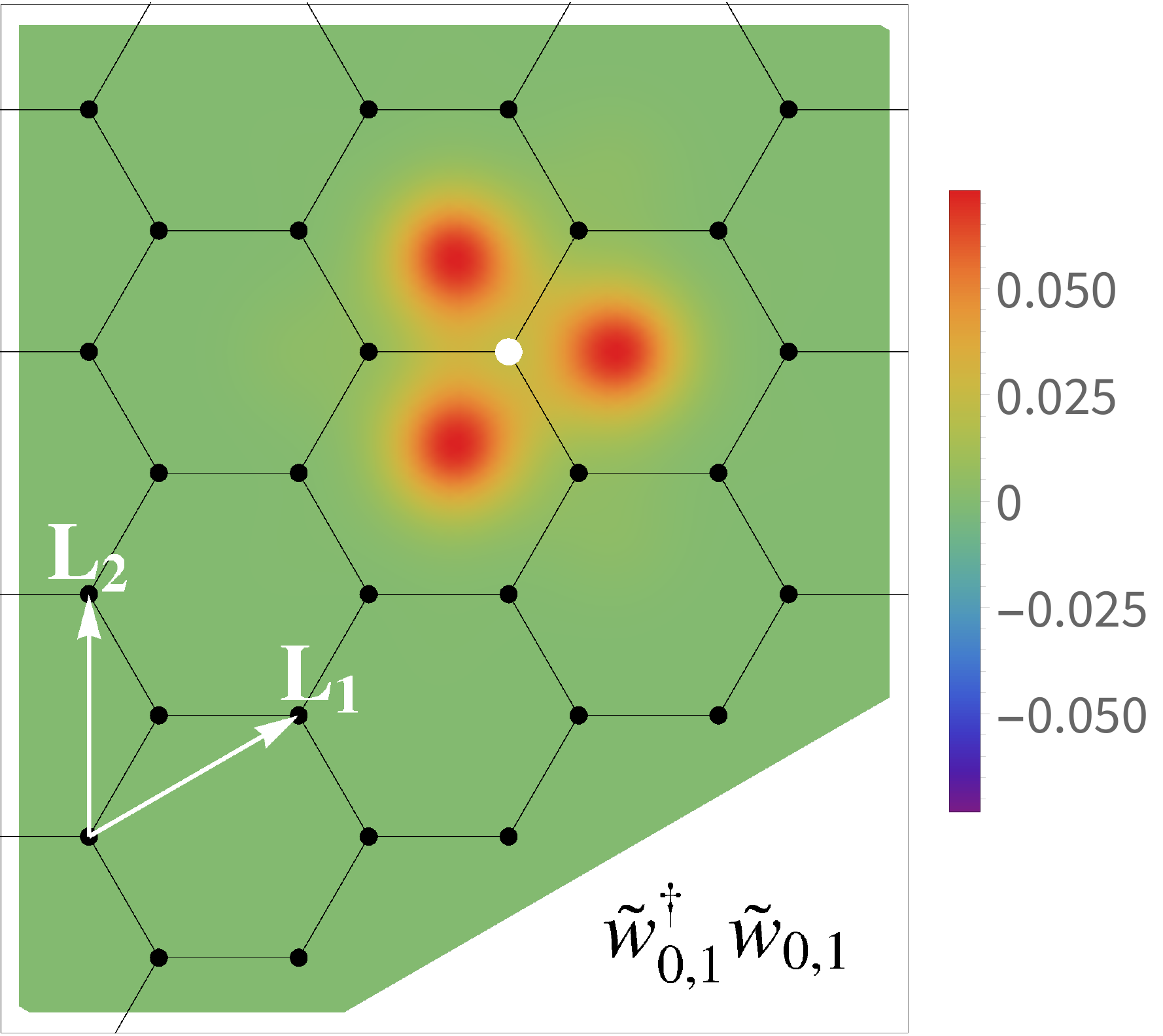}}
	\subfigure[\label{Fig:WannierOverlaps:2}]{\includegraphics[width=0.6\columnwidth]{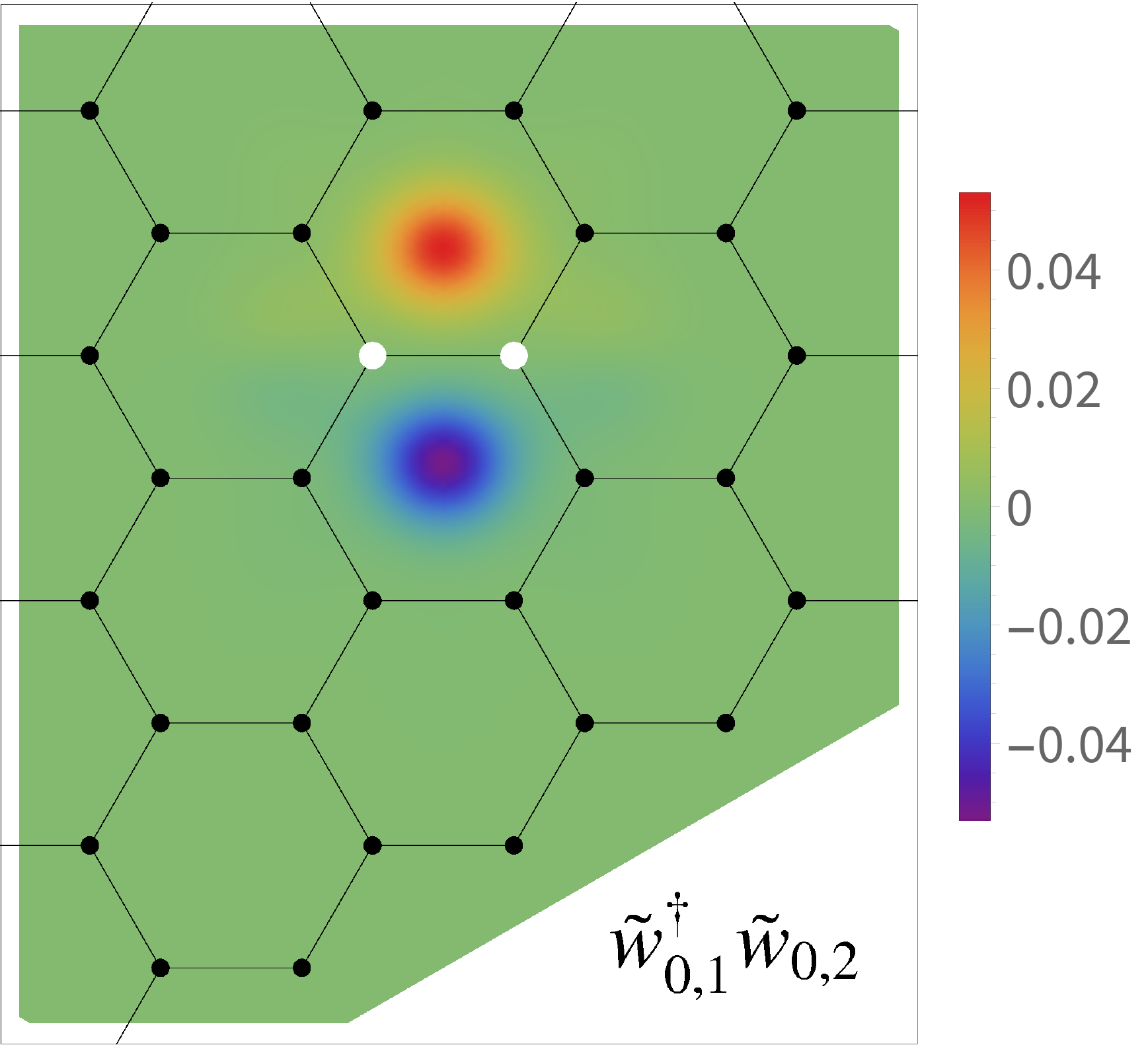}}
	\subfigure[\label{Fig:WannierOverlaps:3}]{\includegraphics[width=0.6\columnwidth]{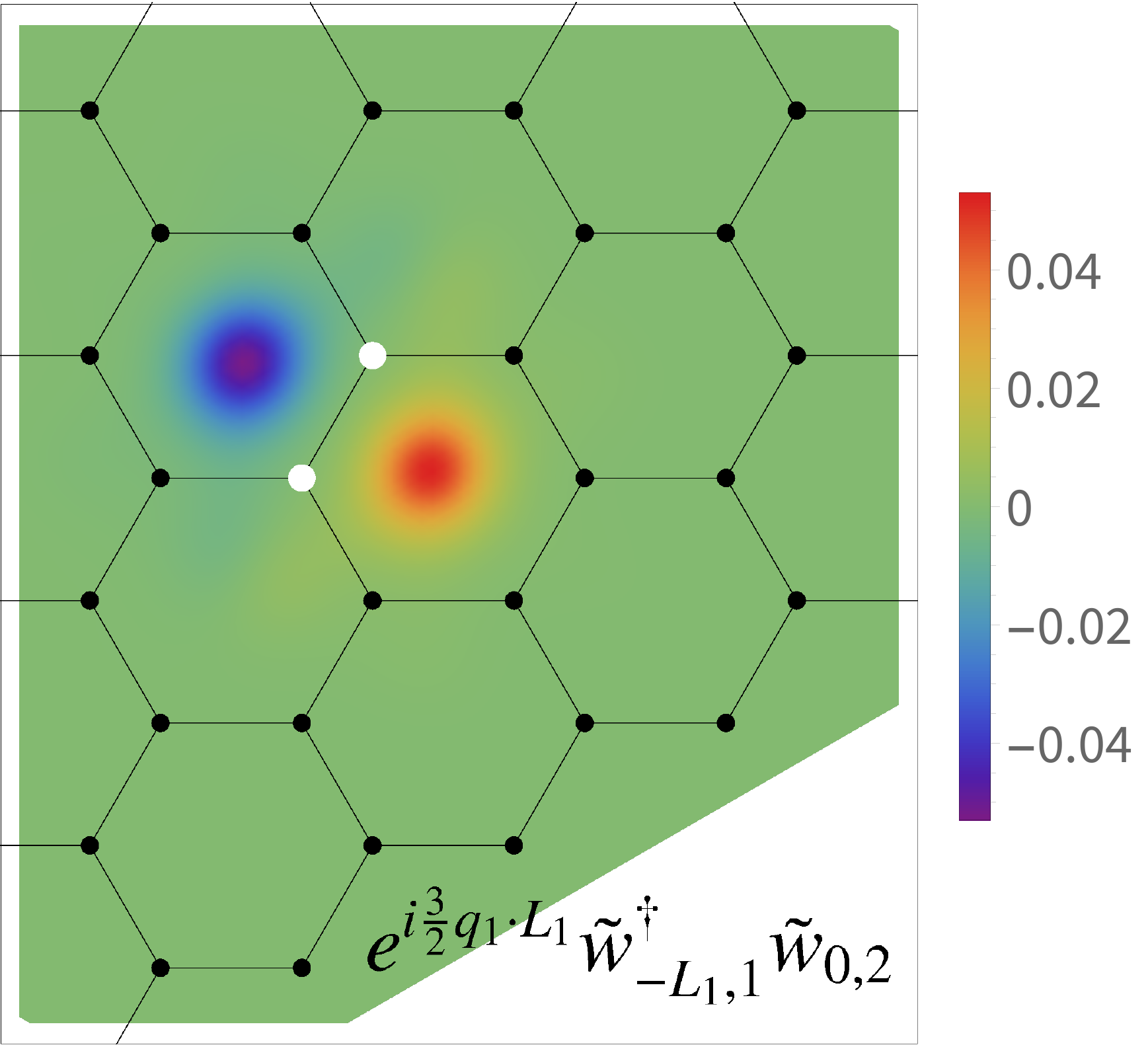}}
	\subfigure[\label{Fig:WannierOverlaps:4}]{\includegraphics[width=0.6\columnwidth]{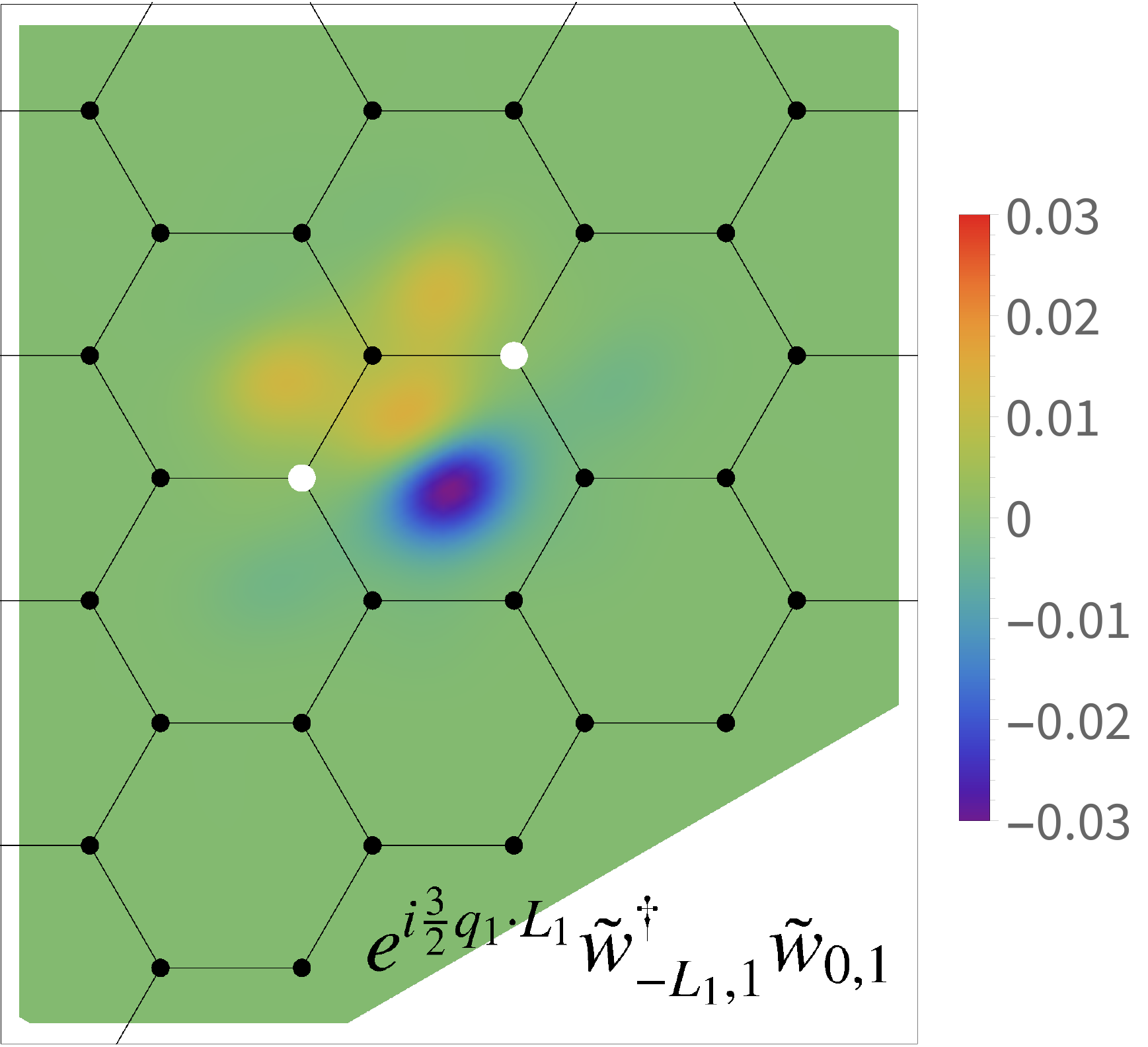}}
	\subfigure[\label{Fig:WannierOverlaps:5}]{\includegraphics[width=0.6\columnwidth]{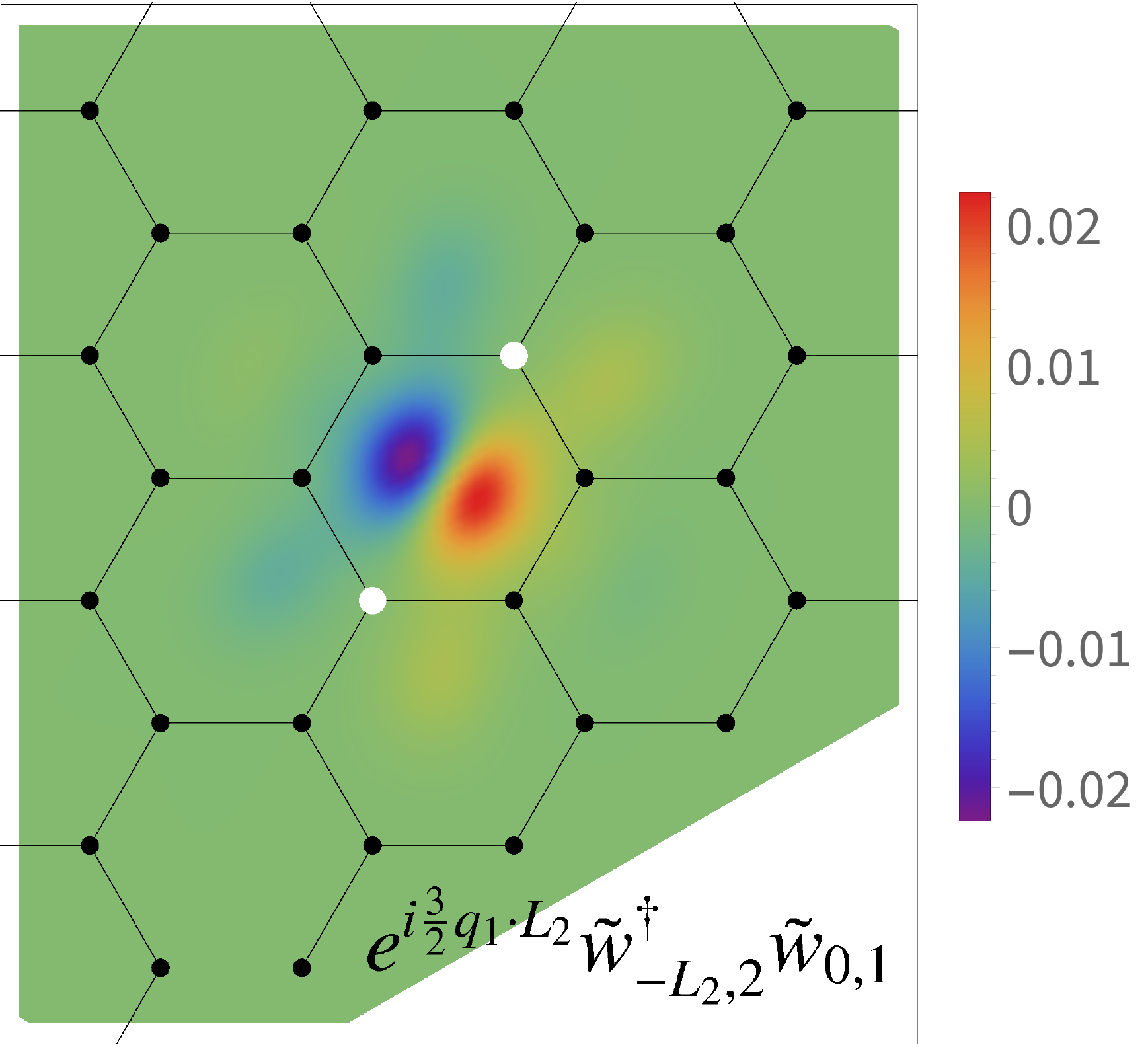}}
	\subfigure[\label{Fig:WannierOverlaps:6}]{\includegraphics[width=0.6\columnwidth]{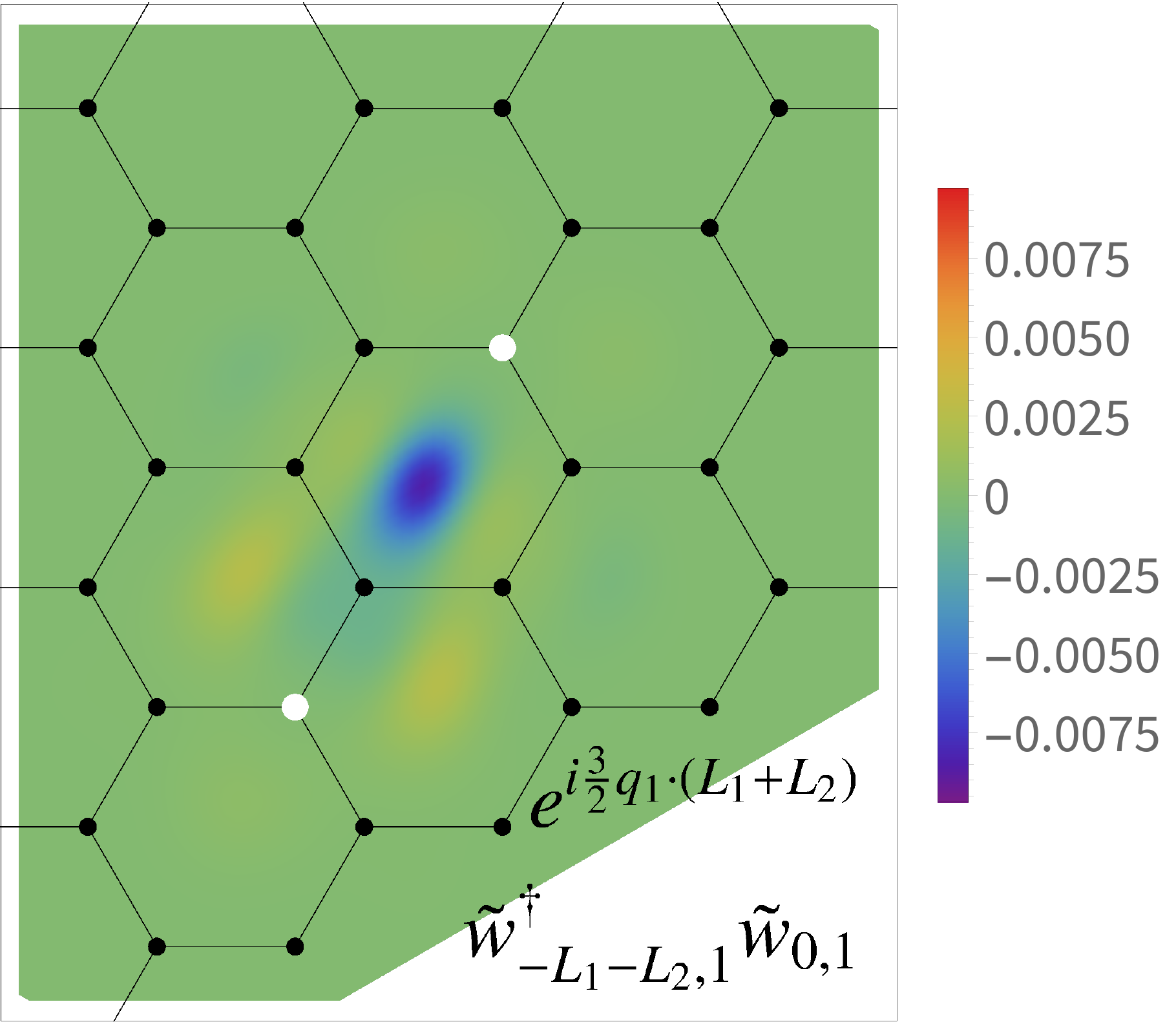}}
	\caption{The real space plot of the combination of the Wannier orbitals entering into the Eqs.~(\ref{eqn:exchangeWtilde}-\ref{eqn:directWtilde}), $e^{-i\frac{3}{2}\bq_1\cdot\bR}\tilde{w}^\dagger_{\bR,j}(\br)\tilde{w}_{\bR',j'}(\br)$, where $\tilde{w}_{\bR,j}(\br)$ is related to $w_{\bR,j}(\br)$ in Eq.(\ref{eqn:wtilde}). As shown in the section IV.A, $P\mathcal{C}C_2T$ symmetry guarantees that these overlap functions are purely real. The Wannier centers, determined by $\bR$ and $j$, are illustrated by the white dots; the panel (a) shows the familiar three peak ``fidget spinner'' structure for on-site overlap (i.e. $\bR=\bR'$ and $j=j'$.) As shown in the panel (d), the overlap for the nearest neighbors on the triangular lattice is predominantly negative near the AA sites. As illustrated in Fig.~\ref{Fig:Hopping:1}, the on-site tight-binding parameter $t_{\mathbf{0}}^{(F)}$ is dominated by the the electrostatic self-interaction of a ``monopole'' (panel (a)), self-interaction of each of the ``dipoles'' shown in panels (b) and (c); it is clearly positive. The nearest neighbor hopping parameter $t_{\mathbf{L}_1}^{(F)}$ is dominated by two terms: the electrostatic interaction between the ``monopole'' (panel (a)) and the ``dipole'' (panel (d)), as well as the electrostatic interaction between the ``dipole'' in panel (b) and the ``dipole'' in panel (c). It is obvious that both of these terms will lead to the negative $t_{\mathbf{L}_1}^{(F)}$. }
	\label{Fig:WannierOverlaps}
\end{figure*}
\end{widetext}

Because these overlaps will be used in determining the form of the projected interactions, it will be helpful to visualize them. This is done in Fig.~\ref{Fig:WannierOverlaps}. 

Finally, note that the exactly flat band dispersion of the kinetic energy at the magic angle of the chiral limit is trivially satisfied, because the action of the kinetic energy operator on the Wannier states annihilates them.

\section{Single particle excitations in the strong coupling in the Wannier state basis}
\begin{figure*}[t]
	\centering
	\subfigure[\label{Fig:AllHopping:Fock}]{\includegraphics[width=0.92\columnwidth]{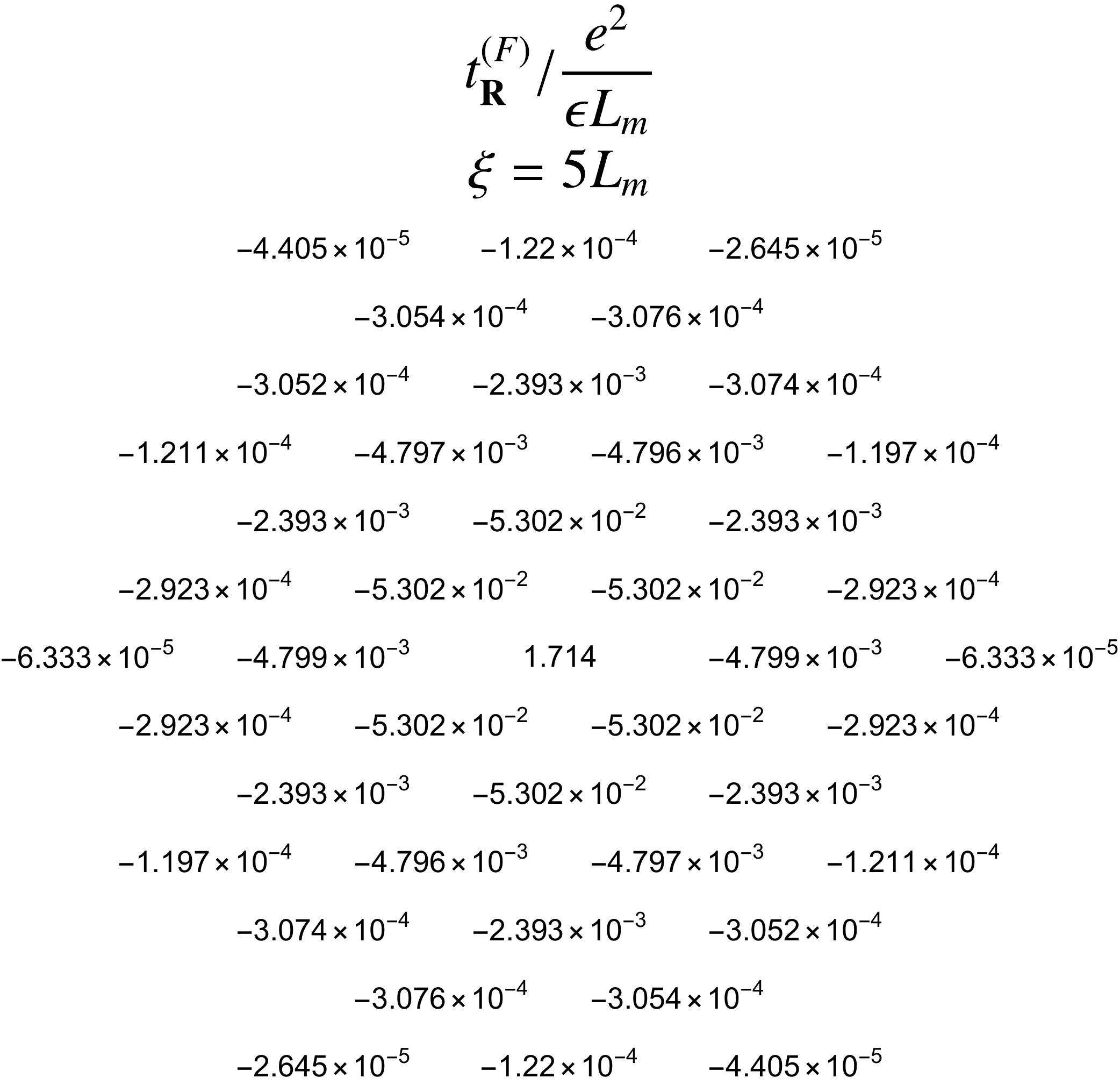}}	\hspace{0.5cm}
	\subfigure[\label{Fig:AllHopping:Hartree}]{\includegraphics[width=0.92\columnwidth]{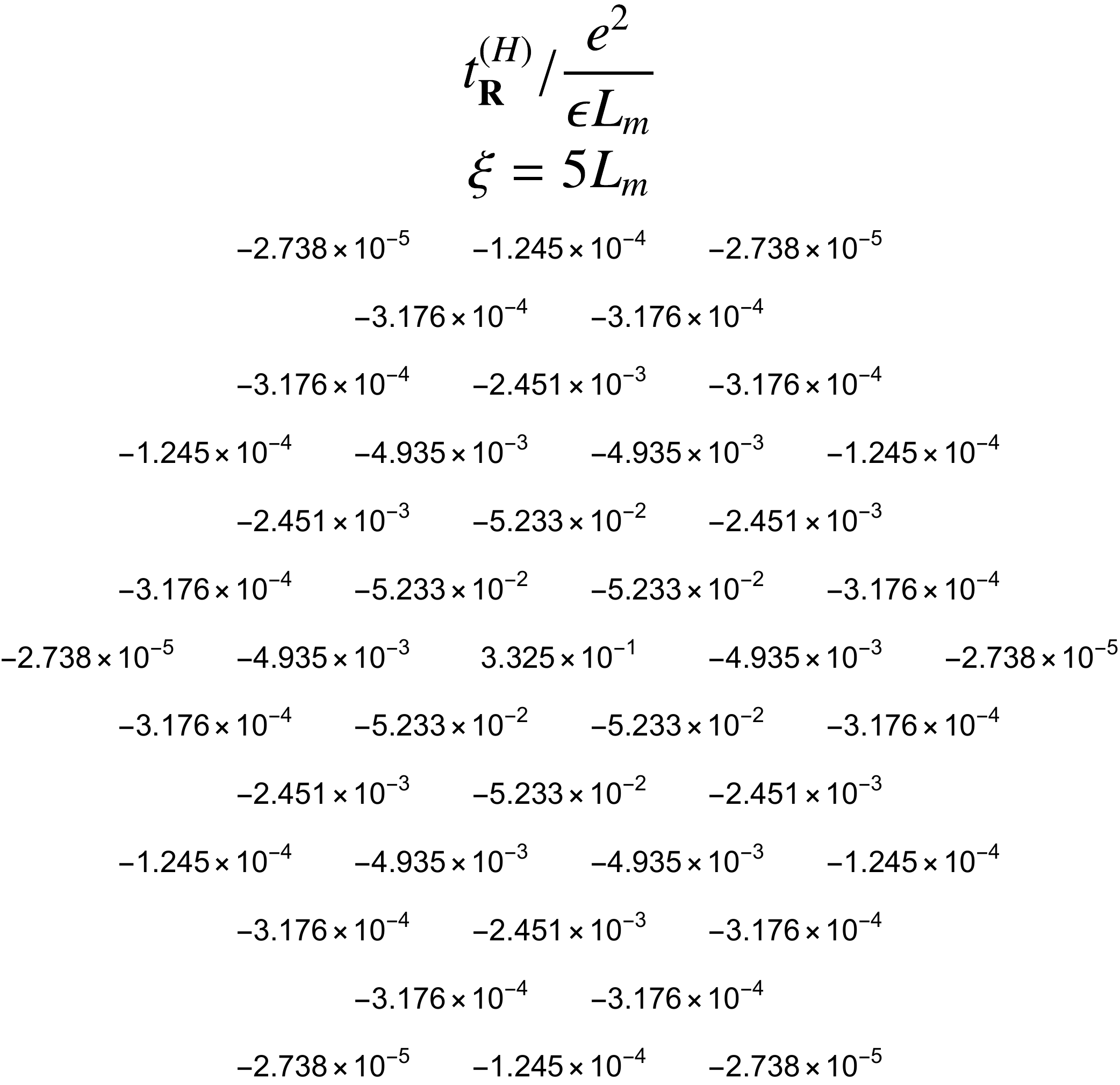}}	
	\caption{The hopping constants, defined in Eqn.~\ref{Eqn:FockHopping} for the exchange term (left) and \ref{Eqn:HartreeHopping} for direct term (right), on the triangular sub-lattice of the honeycomb moire lattice with the screening gate separation $\xi = 5 L_m$.  In both panels, $t_{\mathbf{R}}^{(F/H)}$s are shown at the position of $\mathbf{R}$ relative to the center of the cluster. Note that the hopping constants are negligibly small for $|\mathbf{R}|> 2L_m$ and that the values of the (non-negligible) off-site hopping amplitudes are very similar.}
	\label{Fig:AllHopping}
\end{figure*}

In order to re-write $\mathcal{H}^{h/p}(\bk)$ in the Wannier basis, we first note that
\begin{eqnarray}\label{eqn:Fdef}
&&\mathcal{F}(\br,\br')\equiv \sum_{m\bp}\Psi_{m,\bp}(\br)\Psi^\dagger_{m,\bp}(\br')\\
&&=\sum_{m\bp}\tilde{\Phi}_{j,\bp}(\br)\tilde{\Phi}^\dagger_{j,\bp}(\br')
=\sum_{j,\bR}w_{\bR,j}(\br)w^\dagger_{\bR,j}(\br'),
\end{eqnarray}
where in the first line we used the fact that the transformation from $\Psi_{m,\bp}(\br)$ to $\tilde{\Phi}_{j,\bp}(\br)$ is unitary (as of course is the final transformation to $w$'s).
\begin{widetext}
Using the same sequence of unitary transformations, we can then re-write the Eqn.~\ref{eqn:Hhp} in the Wannier basis as
\begin{eqnarray}\label{eqn:HhpWan}
\mathcal{H}^{h/p}_{nm}(\bk)&\rightarrow& \mathcal{\tilde{H}}^{h/p}_{jj'}(\bk)=\mathcal{E}^{(F)}_{jj'}(\bk)+{\mathcal{E}_{\nu,jj'}^{(H)}}(\bk) \\
&&\mathcal{E}^{(F)}_{jj'}(\bk)=\frac{1}{2N_{uc}}\sum_{\bR\bR'}e^{-i\bk\cdot(\bR-\bR')}\int d^2\br d^2\br' V(\br-\br')
w^\dagger_{\bR,j}(\br)\mathcal{F}(\br,\br')w_{\bR',j'}(\br'),\\
&&{\mathcal{E}_{\nu,jj'}^{(H)}}(\bk)=\mp\frac{\nu}{2N_{uc}}\sum_{\bR\bR'}e^{-i\bk\cdot(\bR-\bR')}
\int d^2\br d^2\br' V(\br-\br') w^\dagger_{\bR,j}(\br')w_{\bR',j'}(\br')\mbox{Tr}\mathcal{F}(\br,\br).
\end{eqnarray}
where $\mathcal{F}(\br,\br')$ was defined in (\ref{eqn:Fdef}). The indices $j$ and $j'$ now label the two sublattices of the moire honeycomb lattice at which the Wannier states are centered.
\end{widetext}
We will now show that $\mathcal{\tilde{H}}^{h/p}_{jj'}(\bk)$ is also proportional to $\delta_{jj'}$ in the chiral limit.
$\mathcal{C}C_2\cT$ guarantees that
\begin{equation}1_2\sigma_y  w^*_{\bR,j}(-\br)
=(-i\lambda_y)_{jj'} w_{-\bR-\bL_2,j'}(\br),\end{equation}
where $\lambda_y=\left(\begin{array}{cc}0 & -i \\
i & 0\end{array}\right)$. In addition, $\mathcal{F}(\br,\br')=1_2\sigma_y \mathcal{F}^*(-\br,-\br')1_2\sigma_y$.
Using these relations and $V(\br)=V(-\br)$ we find that
\begin{eqnarray}
\mathcal{\tilde{H}}^{h/p}_{jj'}(\bk)=(\lambda_y)_{j\ell}\mathcal{\tilde{H}}^{*h/p}_{\ell\ell'}(\bk)(\lambda_y)_{\ell'j''}.
\end{eqnarray}
This means that the $2\times 2$ matrix $\mathcal{\tilde{H}}^{h/p}_{jj'}(\bk)$ must be proportional to $\delta_{jj'}$ for each $\bk$.
Therefore, despite the Wannier centers coinciding with the $AB$ and $BA$ sites forming the moire honeycomb lattice, the spectrum of $\mathcal{\tilde{H}}^{h/p}_{jj'}(\bk)$ and therefore of $\mathcal{H}^{h/p}(\bk)$ can be understood as originating from the hopping on the {\em triangular} moire lattice. In other words, the hopping occurs on the $AB$ triangular sublattice of the honeycomb moire lattice independently of the hopping on the $BA$ triangular sublattice. Because the Wannier states are exponentially localized, the expansion in the range of the hopping is expected to converge fast.

Using (\ref{eqn:w def}) we find that $\mathcal{E}^{(F)}_{jj'}(\bk)$ gives a triangular lattice hopping model with the hopping constants which indeed decay rapidly with distance.
We can thus write $\mathcal{E}^{(F)}_{jj'}(\bk)=\delta_{jj'}\mathcal{E}^{(F)}(\bk)$ where
\begin{eqnarray}
\mathcal{E}^{(F)}(\bk)&=&\sum_{\bR}t^{(F)}_{\bR}e^{i(\bk-\bq_1)\cdot\bR}. \label{Eqn:FockHopping}
\end{eqnarray}
The result for the hopping amplitudes $t^{(F)}_{\bR}$ for the Coulomb interaction is shown in Fig~\ref{Fig:AllHopping:Fock}, using the central site as the reference.
Because the nearest neighbor hopping constant is negative (and because the hopping amplitudes decay rapidly with distance), the minimum of the dispersion is at the $\fvec \Gamma$ point.

For the second (direct) term in (\ref{eqn:HhpWan}), we similarly have $\mathcal{E}^{(H)}_{\nu,jj'}=\mp\delta_{jj'}\mathcal{E}^{(H)}_\nu(\bk)$ where
\begin{eqnarray}
\mathcal{E}^{(H)}_\nu(\bk)&=& \nu \sum_{\bR}t^{(H)}_{\bR}e^{i(\bk-\bq_1)\cdot\bR}, \label{Eqn:HartreeHopping}
\end{eqnarray}
and the hopping amplitudes also decay rapidly with distance as displayed in Fig.~\ref{Fig:AllHopping:Hartree}.

Combining the above results we find that the dispersion
from Eq.(\ref{eqn:Hhp}) is
\begin{eqnarray}
\mathcal{E}^h(\bk)&=&\mathcal{E}^{(F)}(\bk)-  \mathcal{E}^{(H)}_\nu(\bk)\\
\mathcal{E}^p(\bk)&=&\mathcal{E}^{(F)}(\bk)+ \mathcal{E}^{(H)}_\nu(\bk).
\end{eqnarray}

Note that the numerical values of the nearest neighbor hopping constants from the exchange $t^{(F)}_{\bL_{1}}$ and the direct $t^{(H)}_{\bL_{1}}$ are nearly identical.
Therefore, for $\nu=1$ the hole excitations will have a nearly flat dispersion, while the dispersion for the particle steepens by a factor of $\approx 2$.
Similarly, at $\nu=-1$, the flattening happens when a particle is added, and the approximate bandwidth doubling happens when a hole is added. We see that the effective mass of the excitations that bring the filling closer to the charge neutrality point (holes for $\nu=1$ and particles for $\nu=-1$) is much larger than the effective mass for the complementary excitation that move the filling away from the charge neutrality point.

For $\nu=2$ and $\nu=3$, the hole bands are completely inverted due to this effect and the electron bands get steeper~\cite{BABTBGV,KangVafekCascades}. The effective mass is still very large on the hole side.


\section{Dependence of the dispersion on the range of the Coulomb interaction}

\begin{widetext}
\begin{figure*}[t]
	\centering
	\subfigure[\label{Fig:WSMom:1}]{\includegraphics[width=0.6\columnwidth]{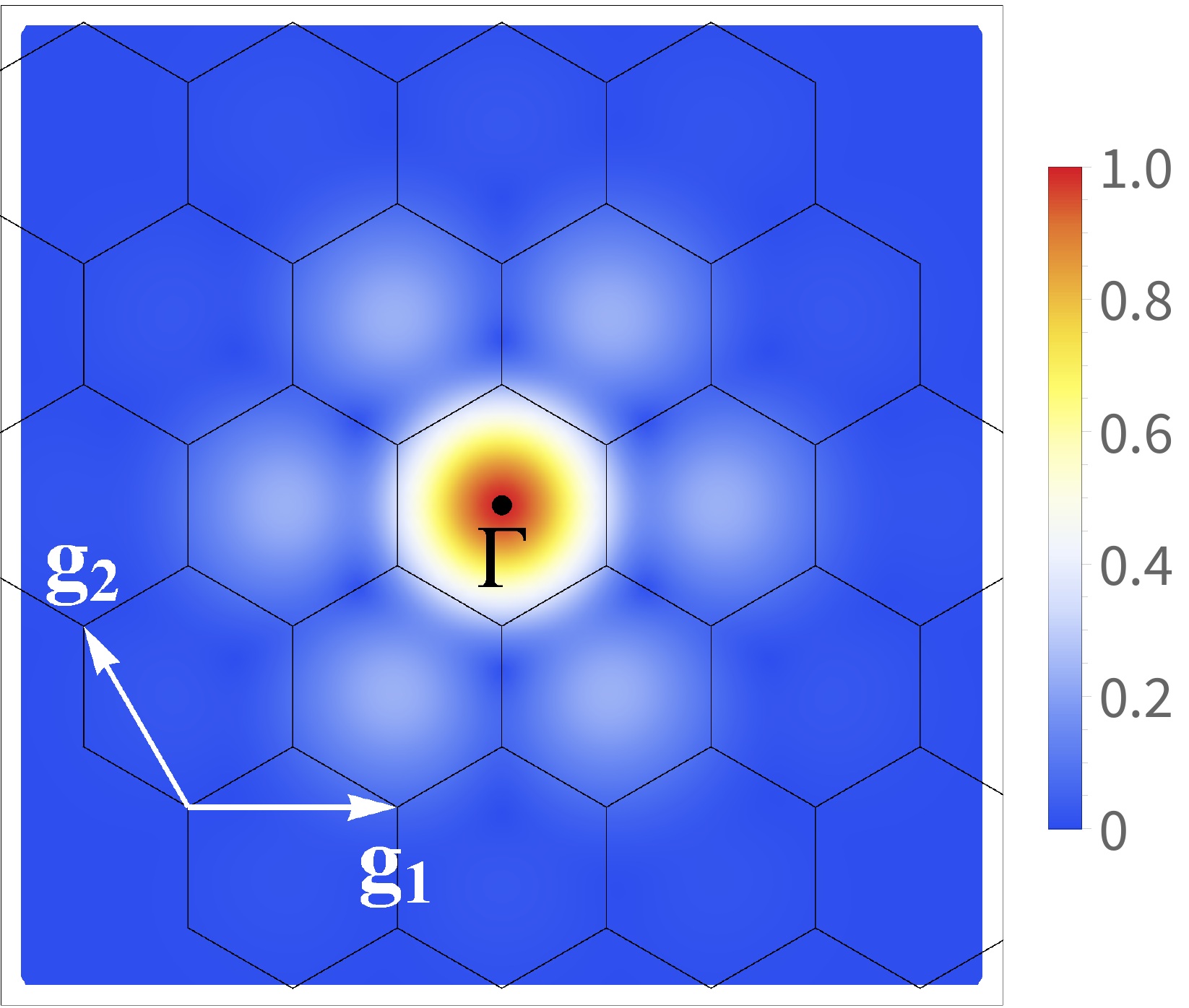}}		\subfigure[\label{Fig:WSMom:2}]{\includegraphics[width=0.6\columnwidth]{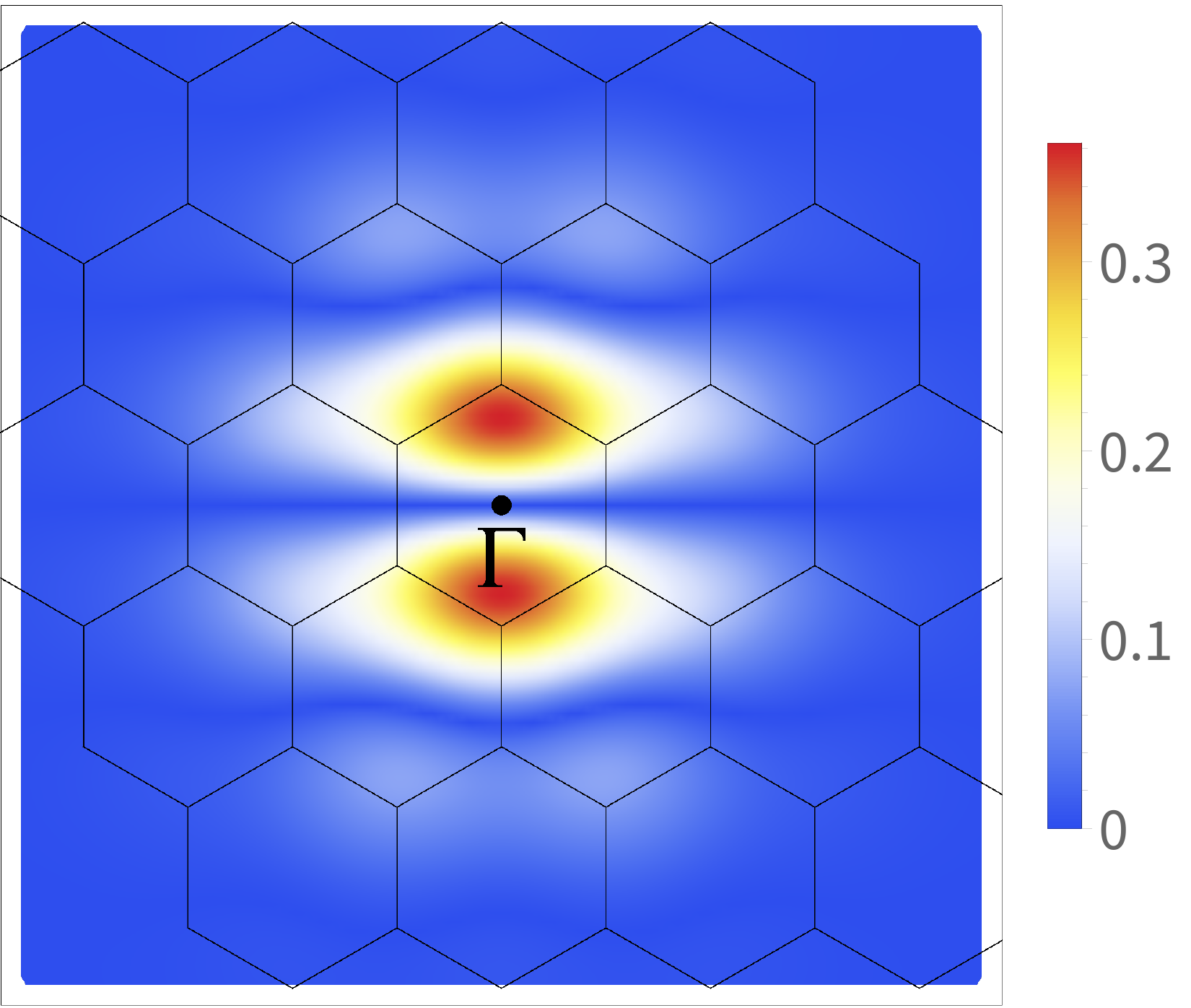}}
	\subfigure[\label{Fig:WSMom:3}]{\includegraphics[width=0.6\columnwidth]{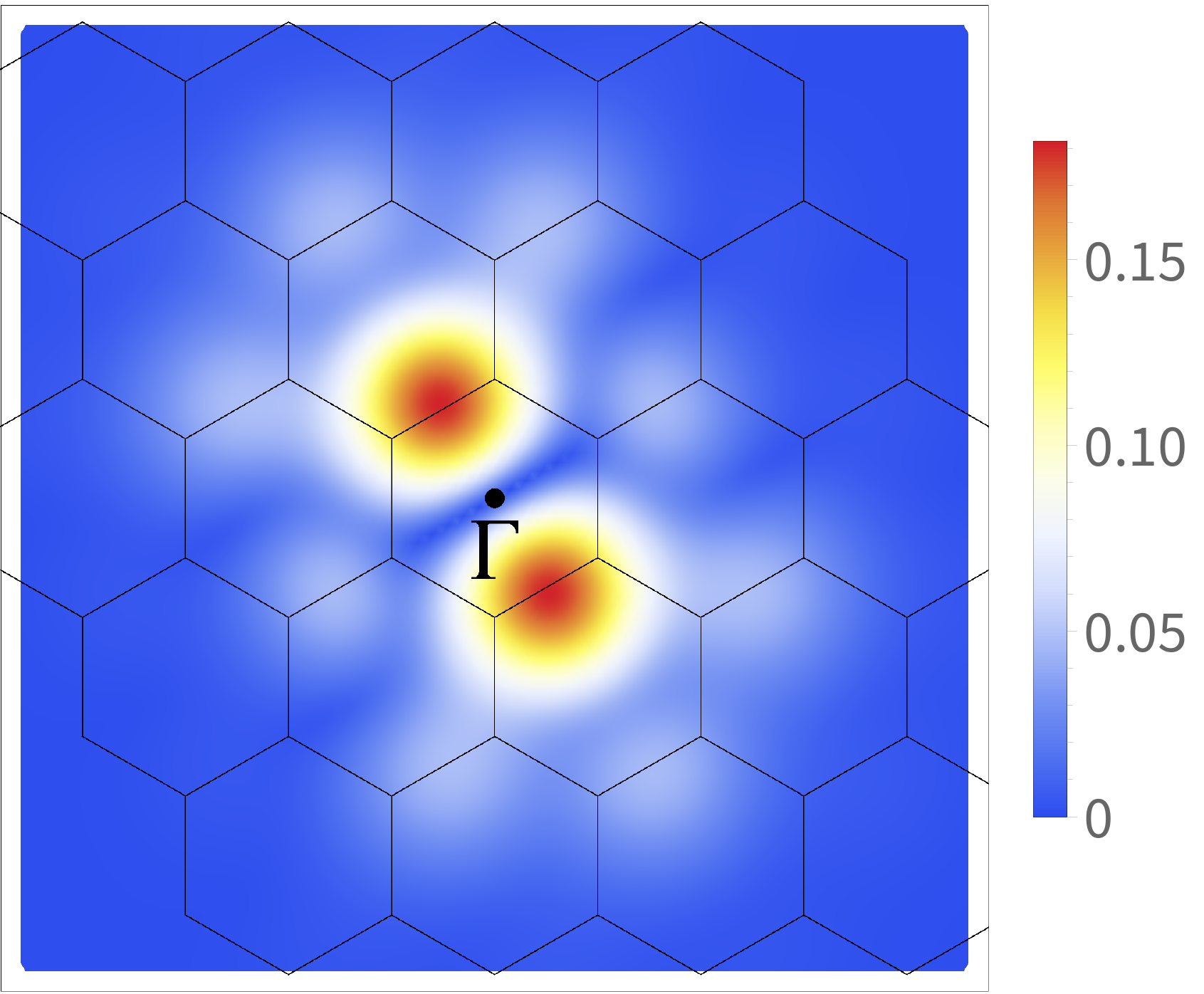}}
	\caption{The plot of $\left|O_{\bR, j; \bR', j'}(\bq)\right|$, where $O_{\bR,j; \bR', j'}(\bq) = \int \rmd^2\br\ \tilde{w}^\dagger_{\bR,j}(\br)\tilde{w}_{\bR',j'}(\br) e^{-i \bq \cdot \br}$ is the Fourier transform of the product of two Wannier orbitals centered at $\bR,j$ and $\bR',j'$. The relation between $\tilde{w}_{\bR,j}(\br)$ and $w_{\bR,j}(\br)$ is given in Eq.(\ref{eqn:wtilde}). (a) The Fourier transform of a ``monopole'' ( $\bR = \bR' = \mathbf{0}$ and $j = j' = 1$) is peaked around $\fvec\Gamma$. When centered on different sites, orthogonality of the Wannier states guarantees that $O_{\bR, j; \bR', j'}(\bq)$ vanishes as $\bq\rightarrow 0$. (b) 
	 ``Dipoles'' ( $\bR = \bR' = \mathbf{0}$, $j= 1$ and $j' = 2$) and (c) ($\bR = - \bL_1$, $\bR'=\fvec{0}$ and $j = j' = 1$) contain two main peaks separated by approximately the primitive reciprocal lattice vector with length $\sim |\bg_{1,2}|$; note that the ``dipoles'' vanish at $\fvec\Gamma$. }
	\label{Fig:WSMom}
\end{figure*}
\end{widetext}

In order to gain a better understanding of our results, it is useful to re-write them in terms of $\tilde{w}$ defined in Eqn.~\ref{eqn:wtilde}. This is partly because then we can take advantage of the Wannier product $\tilde{w}^\dagger_{\bR,j}(\br)\tilde{w}_{\bR',j'}(\br)$ being purely real as shown in Eqn.~\ref{eqn:WS overlap is real}, and partly because our Wannier states are 2D exponentially localized.
\begin{widetext}
To this end, we have
\begin{eqnarray}\label{eqn:exchangeWtilde}
&&\mathcal{E}^{(F)}_{jj'}(\bk)=\sum_{\bR\bR'}e^{-i\left(\bk-\bq_1\right)\cdot(\bR-\bR')}\sum_{i}\frac{1}{2}\int d^2\br d^2\br' V(\br-\br')
\left(e^{-i\frac{3}{2}\bq_1\cdot\bR}\tilde{w}^\dagger_{\bR,j}(\br)\tilde{w}_{\mathbf{0},i}(\br)\right)
\left(\tilde{w}^\dagger_{\mathbf{0},i}(\br') \tilde{w}_{\bR',j'}(\br')e^{i\frac{3}{2}\bq_1\cdot\bR'}\right),\\
&&{\mathcal{E}_{\nu,jj'}^{(H)}}(\bk)=\mp \nu \sum_{\bR\bR'}e^{-i\left(\bk-\bq_1\right)\cdot(\bR-\bR')}
\sum_i \frac{1}{2}\int d^2\br d^2\br' V(\br-\br')\tilde{w}^\dagger_{\mathbf{0},i}(\br)\tilde{w}_{\mathbf{0},i}(\br) \tilde{w}^\dagger_{\bR,j}(\br')\tilde{w}_{\bR',j'}(\br')e^{-i\frac{3}{2}\bq_1\cdot(\bR-\bR')}.
\label{eqn:directWtilde}
\end{eqnarray}
where $\tilde{w}_{\mathbf{0},i}(\br)$ corresponds to the Wannier state centered inside the origin unit cell ($\bR=\mathbf{0}$) either at the $AB$ or $BA$ sites depending on $i$. Note that at the $\fvec\Gamma$ point $\bk=\bq_1$, and the phase factor before the integrals becomes $1$. The phase factors $e^{-i\frac{3}{2}\bq_1\cdot\bR}$ which we intentionally absorbed inside the integrals are actually purely real and equal to $(-1)^{m_1}$ for $\bR=m_1\bL_1+m_2\bL_2$ where $m_{1,2}$ are integers. By comparing with Eqn.~\ref{Eqn:FockHopping} and \ref{Eqn:HartreeHopping}, we obtain the formula for the hopping constants
\begin{align}
    t_{\bR}^{(F)} & = \sum_{\bR'}\frac12 \int\rmd^2 \br \rmd^2 \br'\ V(\br - \br') \sum_i \left( e^{-i\frac32 \bq_1 \cdot \bR'} \tilde{w}^{\dagger}_{\bR', 1}(\br) \tilde{w}_{\mathbf{0}, i}(\br) \right) \left( \tilde{w}^{\dagger}_{\mathbf{0}, i}(\br') \tilde{w}_{\bR' + \bR, 1}(\br') e^{i\frac32 \bq_1 \cdot (\bR' + \bR)} \right), \label{Eqn:FockHoppingExp}  \\
    t_{\bR}^{(H)} & = e^{i\frac32 \bq_1 \cdot \bR}  \sum_{\bR'} \frac12 \int\rmd^2 \br \rmd^2 \br'\ V(\br - \br') \sum_i \left( \tilde{w}^{\dagger}_{\mathbf{0}, i}(\br) \tilde{w}_{\mathbf{0}, i}(\br) \right) \left( \tilde{w}^{\dagger}_{\bR', 1}(\br') \tilde{w}_{\bR' + \bR, 1}(\br') \right).   \label{Eqn:HartreeHoppingExp}
\end{align}
\end{widetext}

\begin{widetext}
\begin{figure*}[t]
	\centering
	\subfigure[\label{Fig:Hopping:1}]{\includegraphics[width=1.2\columnwidth]{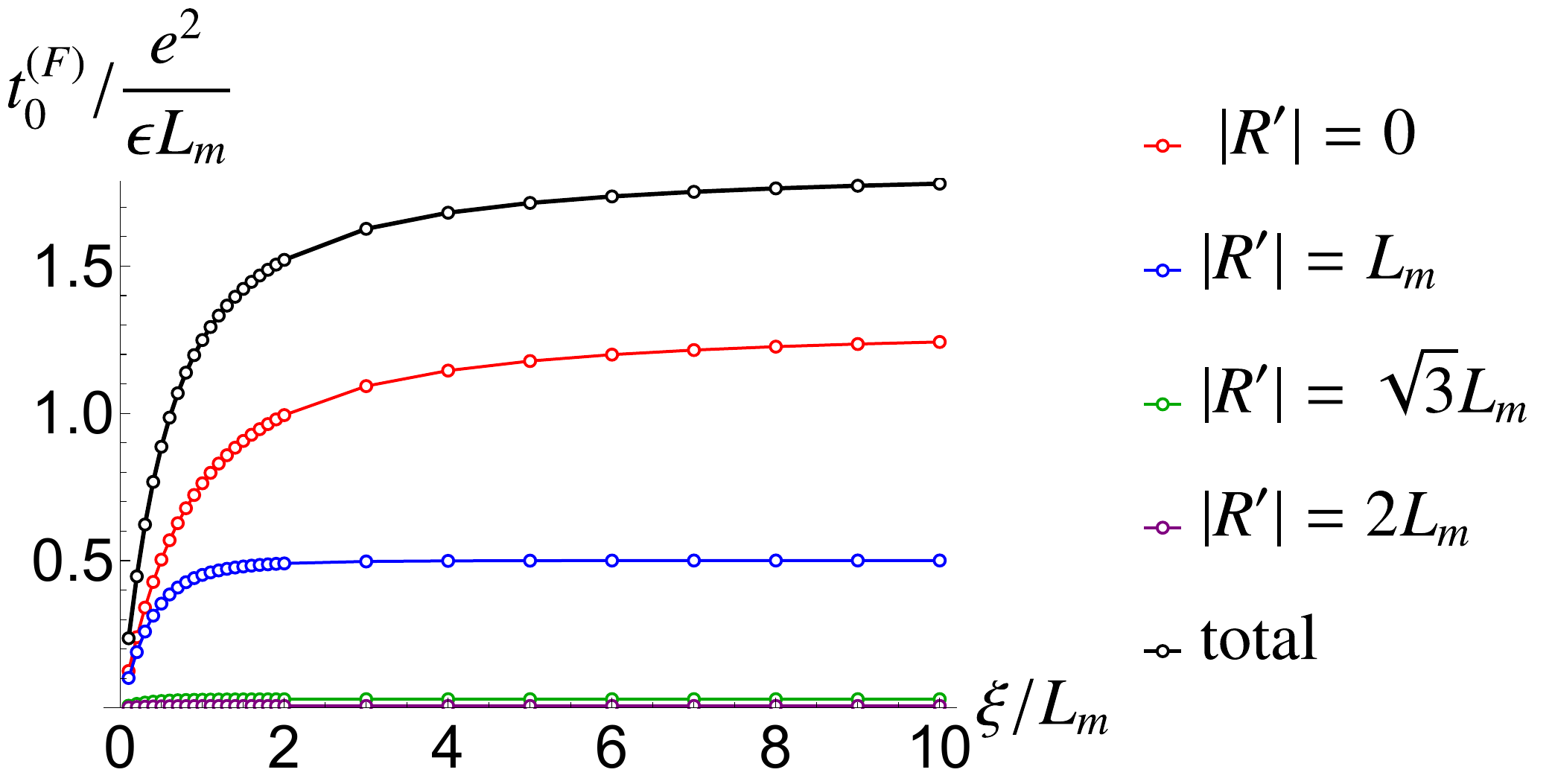}}	\hspace{0.5cm}
	\subfigure[\label{Fig:Hopping:2}]{\includegraphics[width=0.5\columnwidth]{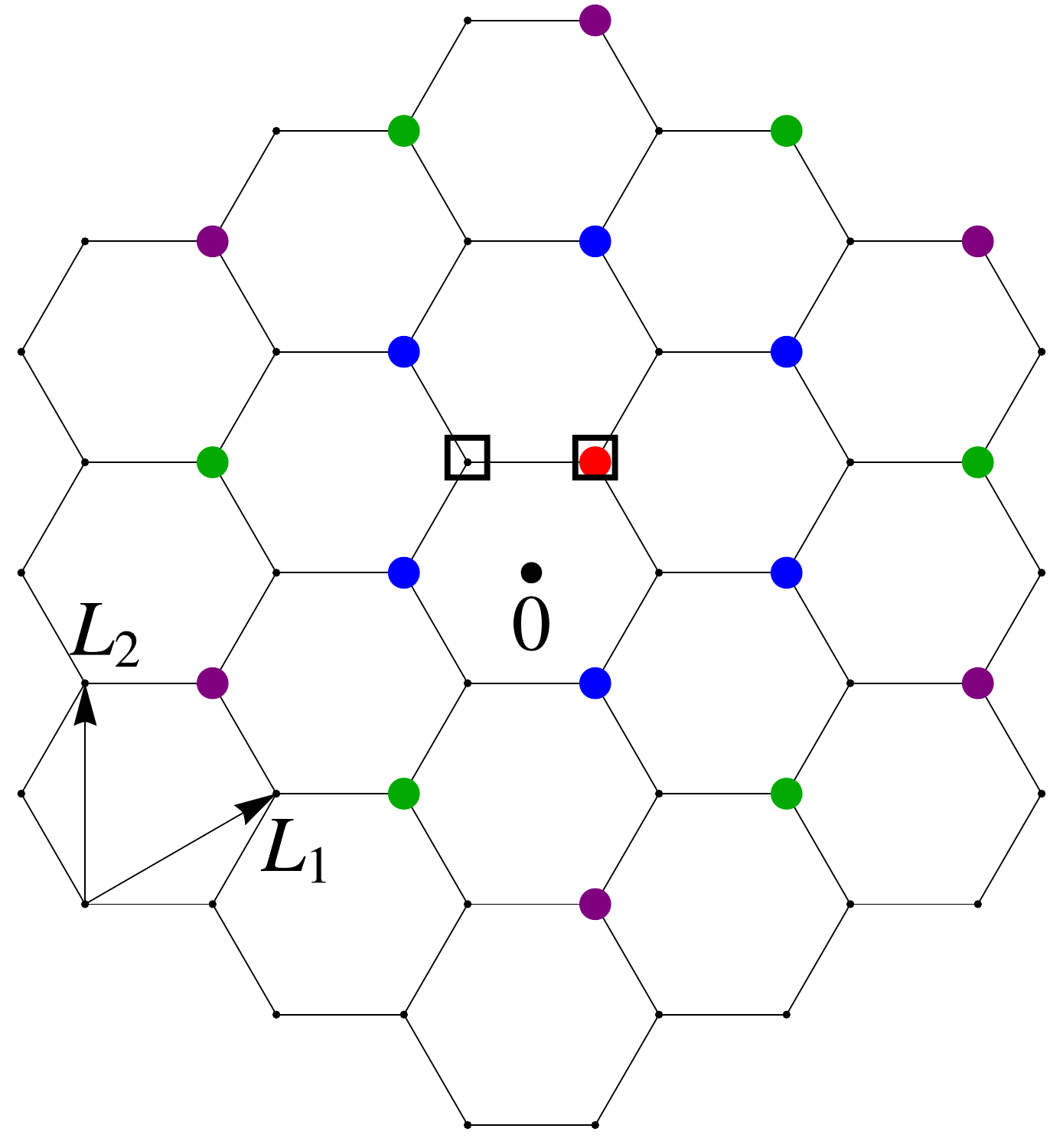}}	
	\subfigure[\label{Fig:Hopping:3}]{\includegraphics[width=1.4\columnwidth]{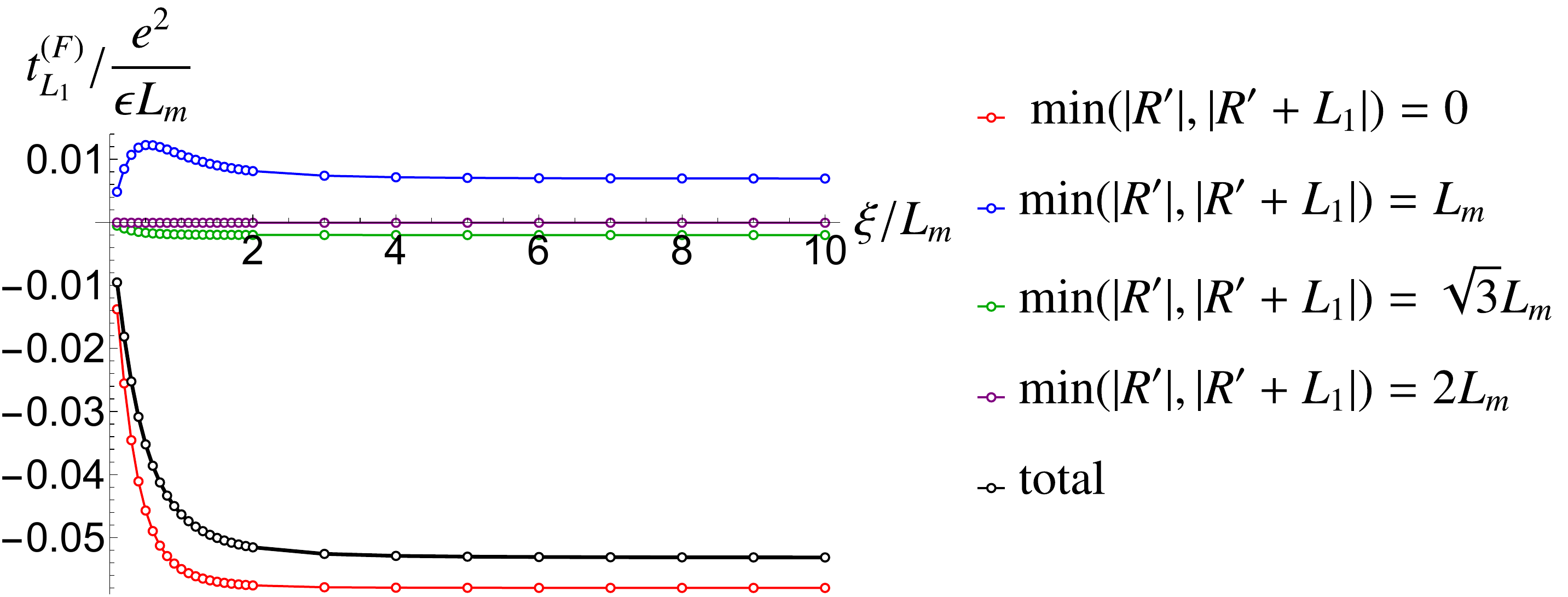}} \hspace{0.5cm}
	\subfigure[\label{Fig:Hopping:4}]{\includegraphics[width=0.5\columnwidth]{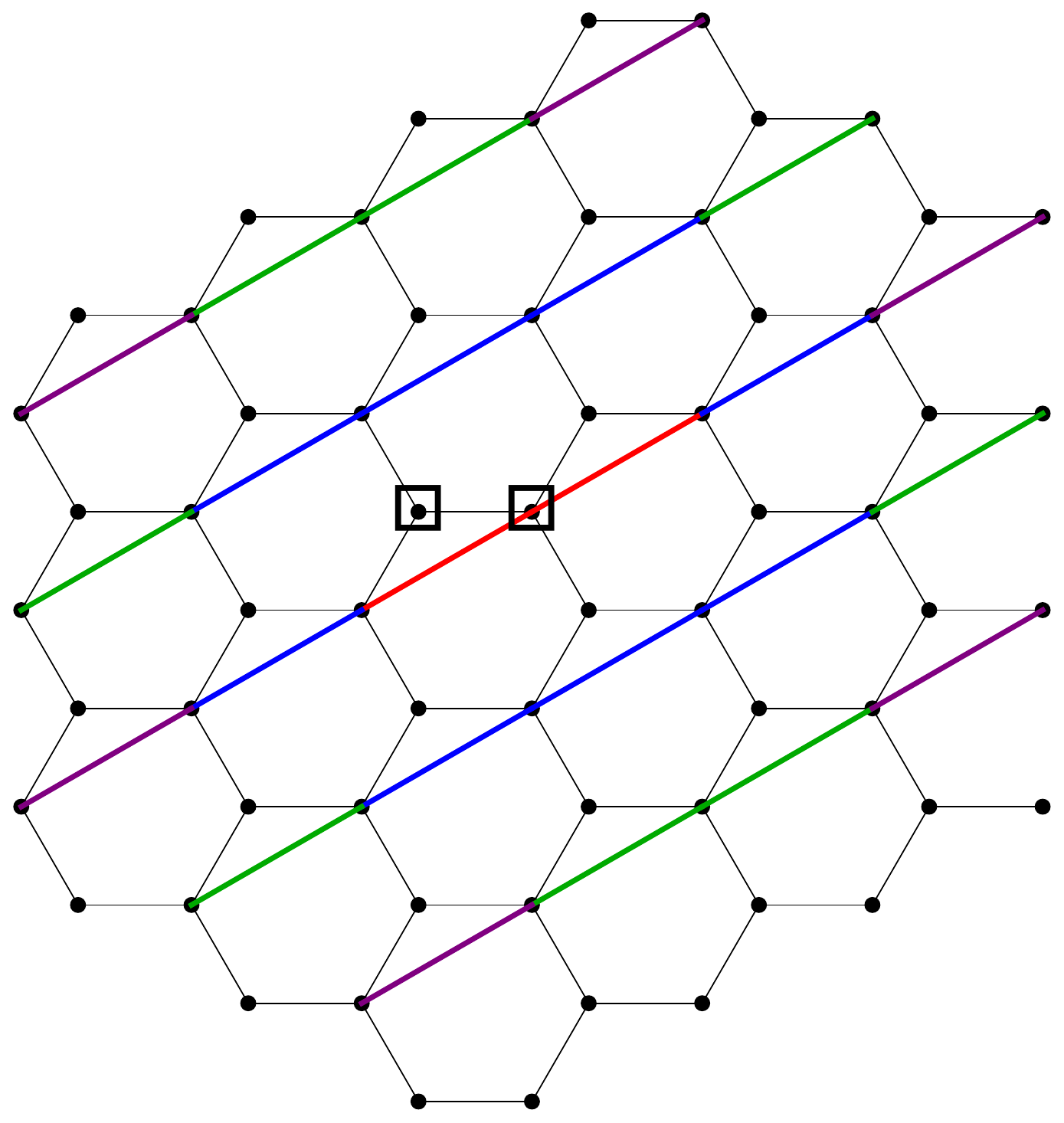}}
	\caption{The exchange contributions to the  on-site (above) and the nearest-neighbor hopping (below) terms of the tight binding model on the triangular lattice. Both the on-site ($\bR=\mathbf{0}$) and the nearest neighbor ($\bR=\bL_1$) hopping parameters are obtained by summing the contributions from different $\bR'$ in the Eqn.~\ref{Eqn:FockHoppingExp}.  The black curve in the panel (a) shows $t^{(F)}_{\bR=\mathbf{0}}$ resulting from the sum over all $\bR'$ for different values of $\xi$. The other colored curves plot the sum over a restricted set of $\bR'$s illustrated in the panel (b) using the same color. The two open squares in the panel (b) illustrate the term $\tilde{w}_{\mathbf{0},i}(\br)\tilde{w}^\dagger_{\mathbf{0},i}(\br')$ appearing in the Eqn.~\ref{Eqn:FockHoppingExp} for $i=1$ (right open square) and $i=2$ (left open square). Each term contributing to the sum over $\bR'$ and the sum over $i$ for $\bR=\mathbf{0}$ is then represented by connecting one of the open squares to one of the colored solid dots in the panel (b), with the corresponding charge distributions illustrated in the Fig.~\ref{Fig:WannierOverlaps}.
	Similarly, the colored curves of $t^{(F)}_{\bR=\bL_1}$ in the panel (c) correspond to those contributions to the sum over $\bR'$ and $i$ in the Eqn.~\ref{Eqn:FockHoppingExp} that are obtained by connecting one of the open squares to the two ends of each colored bond $(\bR', \bR'+ \bL_1)$ in the panel (d).} 
	\label{Fig:Hopping}
\end{figure*}
\end{widetext}

The key insight is to think about each term contributing to the hopping amplitude on a particular bond separately as an electrostatics problem with different charge density distributions. In this way, let us first consider the contribution to $\mathcal{E}^{(F)}_{jj'}(\bk)$ from $\bR=\bR'$ (we already established that $j=j'$ is the only non-zero contribution, so we will assume this implicitly). This is the exchange contribution to the on-site term in our effective tight-binding description of the strong coupling single particle dispersion.
Thus, within the sum over all terms for which $\bR=\bR'$, we have one term when $\bR=\bR'=0$ and $j=i$ which is equivalent to the electrostatic energy of two coinciding, and 2D localized, ``fidget spinners'' (see Fig~\ref{Fig:WannierOverlaps:1}). The multipole expansion of the (purely real) charge distribution $\tilde{w}^\dagger_{0,i}(\br)\tilde{w}_{0,i}(\br)$ therefore contains a monopole term. In momentum space, this contribution then has the form $\int d^2\bq V_\bq \rho_m(\bq)\rho_m(-\bq)$. Because $\rho_m(\bq)$ is peaked at small $\bq$ (see Fig.~\ref{Fig:WSMom:1}), this integral will be sensitive to the small $\bq$ behaviour of $V_\bq$. This is indeed what we see in Fig.~\ref{Fig:Hopping:1} where we show the dependence of this contribution on the distance to the screening gates $\xi$, symmetrically positioned above and below the twisted bilayer graphene in which case $V_\bq=\frac{2\pi e^2}{\eps q}\tanh\frac{q\xi}{2}$ for $\bq\neq 0$, otherwise at $\bq=0$ it vanishes. Note that even when the gates are $\sim 5L_m$ away from the twisted bilayer, the onsite term is still visibly $\xi$-dependent.
The remaining contribution to the on-site $\bR=\bR'$ term comes from either $\bR=\bR'\neq 0$ or $i\neq j$. Our electrostatics problem now deals with charge distributions for which the two Wannier states are not centered on the same site, which means they are orthogonal when
$\tilde{w}^\dagger_{\bR,j}(\br)\tilde{w}_{0,i}(\br)$ is integrated over all $\br$; note that 
$e^{-i\frac32\bq_1\cdot\bR}\tilde{w}^\dagger_{\bR,j}(\br)\tilde{w}_{0,i}(\br)$ is also purely real for any $\br$. This means that the monopole contribution must be absent.
As shown in the Fig.~\ref{Fig:WannierOverlaps}, the distributions $e^{-i\frac32 \bq_1 \cdot\bR} \tilde{w}^\dagger_{\bR,j}(\br)\tilde{w}_{0,i}(\br)$ indeed look dipolar. In Fourier space, $\rho_d(\bq)$ must therefore vanish as $\bq\rightarrow 0$ linearly in $\bq\cdot \bd$ where $\bd$ is the direction of the dipole moment. As a result, the contribution from $\int d^2\bq V_\bq \rho_d(\bq)\rho_d(-\bq)$ is much less sensitive to the small $\bq$ behavior of $V_\bq$. This is indeed seen in Fig.~\ref{Fig:Hopping}, where the dependence on $\xi$ from these contributions saturates at a significantly smaller $\xi$ than for the monopole contributions. Note that because our Wannier states are exponentially localized, the contribution to the on-site term in our tight-binding expansion from Wannier states separated by more than $\sqrt{3}L_m$ is negligibly small, and therefore only bonds in the vicinity of the central site need to be considered.

The contribution to the $\bR\neq \bR'$ hopping terms in our tight-binding description must therefore contain at least one non-monopole distribution. All such contributions are significantly less sensitive to the small $\bq$ behaviour of $V_\bq$ than the monopole-monopole term. This translates to the quicker saturation of the hopping constants with increasing $\xi$ as shown in the Fig.~\ref{Fig:Hopping:3}. 
In addition, visually inspecting the charge distribution contributing to the nearest neighbor hopping displayed in Fig.~\ref{Fig:WannierOverlaps} we see that the contributions coming from the mononpole-dipole terms and the nearest neighbor bond dipole-dipole terms, are clearly negative. Since these contributions dominate, this explains why the minimum of the triangular lattice dispersion is at the $\fvec \Gamma$ point. Further hopping constants are also clearly falling off fast with the range of the hopping.

The contribution of the direct term ${\mathcal{E}_{\nu,jj'}^{(H)}}(\bk)$ can also be understood using this electrostatics analogy. The sum over $\bR$ and $\bR'$, while holding $\bR-\bR'$ fixed, of $\tilde{w}^\dagger_{\bR,j}(\br')\tilde{w}_{\bR',j'}(\br')$ leads to a charge distribution which is periodic in space. Therefore, only reciprocal lattice momenta $\bg$ contribute to the electrostatic energy. The $\bq=0$ term vanishes because $V_{\bq=0}$ vanishes, and the contributions start from $|\bq|=|\bg_{1,2}|=\frac{4\pi}{\sqrt{3}L_m}$. This probes length scales which are shorter than $L_m$, which explains the relative insensitivity of the contribution to the tight-binding amplitudes from the direct term shown in the Fig.~\ref{Fig:TBM:Hartree}.

We therefore reach an interesting conclusion: the strong coupling single particle dispersion in the chiral limit is well approximated by the triangular lattice hopping model, with hopping amplitudes which decay rapidly  with distance past $2L_m$. The excitation gap, being sensitive to the on-site monopole-monopole term in the exchange contribution grows with the range of the Coulomb interactions even when $\xi\gtrsim 5L_m$; this is consistent with the recent experiment in which the excitation gap becomes larger as the small momentum part of the interaction $V_{\bq}$ increases~\cite{LiScience2021}.  On the other hand, the remaining hopping constants are largely insensitive to the range for $\xi\gtrsim L_m$, implying that the effective mass depends on the range of the interaction much more weakly for realistic placement of the screening gates.

\section{Summary}
In this paper we provided a detailed analysis of the strong coupling dispersion in the chiral limit $w_0/w_1=0$. We did so first in the Bloch basis and then using 2D exponentially localized Wannier states which transform simply under valley $U(1)$ symmetry, time reversal symmetry, the unitary particle-hole symmetry $P$, the combination of the chiral particle-hole symmetry, 2-fold rotation about the axis perpendicular to the plane of the twisted bilayer and time reversal ${\mathcal C}C_2\cT$, as well as $C'_2\cT$ (in-plane 2-fold axis shown in the Fig.~\ref{Fig:Schematic:Lattice} followed by the time reversal) and three-fold rotation about the perpendicular $C_3$ symmetries. The $C_2\cT$ symmetry remains obstructed; this should not be taken to mean that it is absent, given that the transformation from Bloch to 2D exponentially localized Wannier states is perfectly unitary. Rather, the $C_2\cT$ symmetry is not explicit. It can in principle be recovered exponentially fast along the lines outlined in the Ref.~\cite{WangPRB2020}.

The 2D exponentially localized Wannier description allows us to understand why the strong coupling single particle dispersion in the chiral limit is well described by the nearest neighbor hopping problem on the {\it triangular} lattice, with a gate distance sensitive on-site term. The negative nearest neighbor hopping term, which is expected from the real space shape of the overlaps shown in the Fig.~\ref{Fig:WannierOverlaps}, explains why the minimum of this spectrum is at the $\fvec \Gamma$ point. These features are not easy to understand directly in the Bloch basis. 

This work therefore highlights not only the ability to in principle construct the 2D localized Wannier basis for the chiral limit of the magic angle twisted bilayer graphene narrow bands, but also their practical utility in understanding non-trivial features of the strong coupling excitation spectra. 
Thus, complementing the Bloch basis calculations with the Wannier basis analysis provides a more powerful way to understand the complexity of the correlated electrons in topologically non-trivial narrow bands.

\acknowledgments
We would like to thank Prof. B. Andrei Bernevig for valuable discussions. O.~V.~is supported by NSF DMR-1916958 and partially by the National High Magnetic Field Laboratory through NSF Grant No.~DMR-1157490 and the State of Florida.
J.~K.~is supported by  NSFC Grant
No.~12074276 and Priority Academic Program Development (PAPD) of Jiangsu Higher Education Institutions.

\newpage
\begin{widetext}
\appendix
\section{Detailed derivation of the symmetries of the Wannier states}
\label{sec:ApxWan}
The exponentially localized Wannier states are
\begin{eqnarray}\label{eqn:Apx wDef}
w_{\bR,j}(\br)&=& \frac{1}{\sqrt{N_{uc}}}\sum_{m,\bk} e^{-i\bk\cdot\bR} \Phi_{m,\bk}(\br)\text{UP}\left[\int d^2\br' \Phi^\dagger_{m,\bk}(\br')h_{j}(\br')\right]
\end{eqnarray}
where the triangular moire lattice is spanned by $\bR=m_1\bL_1+m_2\bL_2$ with integer $m_{1,2}$, and UP stands for the ``unitary part'' of the $2\times 2$ matrix $A_{mj}(\bk)$ (Eqn.\ref{eqn:overlapmatrixA}).
As defined in the main text,
\begin{eqnarray}
h_1(\br)&=&\left(\begin{array}{c} i \\ 0 \\ 0 \\ -\veps^*
\end{array}\right)\delta\left(\br-\br_{AB}\right),\;\;\;\;
h_2(\br)=\veps^* C'_2T h_1(\br)=\left(\begin{array}{c} 0 \\ -1 \\ -i\veps^* \\ 0
\end{array}\right)\delta\left(\br-\br_{BA}\right),
\end{eqnarray}
where $\br_{AB}=\frac{1}{3}\left(\bL_1+\bL_2\right)$, $\br_{BA}=\frac{1}{3}\left(2\bL_2-\bL_1\right)$, $\veps=e^{2\pi i/3}=e^{-i\bq_1\cdot\br_{AB}}=e^{-i\bq_1\cdot\br_{BA}}$, and $\delta(\br)$ is the 2D Dirac $\delta$-function.

\subsection{Action of the symmetry operation $P$}
This symmetry operation was discussed in detail in Section~\ref{sec:Symmetries Bloch}.
Its action on the Wannier state is
\begin{eqnarray}
-i\mu_y w_{\bR,j}(-\br)&=&
\frac{1}{\sqrt{N_{uc}}}\sum_{m,\bk} e^{-i\bk\cdot\bR} \left(-i \mu_y\right)\Phi_{m,\bk}(-\br)\text{UP}\left[\int d^2\br' \Phi^\dagger_{m,\bk}(\br')h_{j}(\br')\right]\\
\label{eqn:ApxP a}
&=&
\frac{1}{\sqrt{N_{uc}}}\sum_{m,\bk} e^{-i\bk\cdot\bR} \left(-i\mu_y\right)\Phi_{m,\bk}(-\br)\text{UP}\left[\int d^2\br' \Phi^\dagger_{m,\bk}(-\br') i\mu_y\left(-i\mu_y\right)h_{j}(-\br')\right]\\
\label{eqn:ApxP b}
&=&
\frac{1}{\sqrt{N_{uc}}}\sum_{m,\bk} e^{-i\bk\cdot\bR} \Phi_{m,-\bk-\bq_1}(\br)\text{UP}\left[\int d^2\br' \Phi^\dagger_{m,-\bk-\bq_1}(\br') \left(-i\mu_y\right)h_{j}(-\br')\right]\\
&=&
\frac{1}{\sqrt{N_{uc}}}\sum_{m,\bk} e^{i(\bk+\bq_1)\cdot\bR} \Phi_{m,\bk}(\br)\text{UP}\left[\int d^2\br' \Phi^\dagger_{m,\bk}(\br') \left(-i\mu_y\right)h_{j}(-\br')\right].
\end{eqnarray}
Going from (\ref{eqn:ApxP a}) to (\ref{eqn:ApxP b}) we used Eqn.\ref{eqn:PforPhi} and the fact that any $\bk$-dependent phase factors acquired by $\Phi$ are canceled by those coming from $\Phi^\dagger$ because they can be taken outside the UP (as is readily seen when considering SVD).
Substituting our trial function (\ref{eqn:h1}) and (\ref{eqn:h2}) we have
\begin{eqnarray}
\left(-i\mu_y\right)h_{1}(-\br')&=&
\left(\begin{array}{c} 0 \\ \veps^* \\ i \\ 0
\end{array}\right)\delta\left(\br+\br_{AB}\right)=\left(\begin{array}{cccc}
-\veps^* & 0 & 0 & 0\\
0 & -\veps^*  & 0 & 0\\
0 & 0 & -\veps & 0\\
0 & 0 & 0 & -\veps
\end{array}\right)h_2\left(\br+\bL_2\right)
,\\
\left(-i\mu_y\right)h_{2}(-\br')&=&\left(\begin{array}{c} i\veps^* \\ 0 \\ 0 \\ -1
\end{array}\right)\delta\left(\br+\br_{BA}\right)=
\left(\begin{array}{cccc}
\veps^* & 0 & 0 & 0\\
0 & \veps^* & 0 & 0 \\
0 & 0 & \veps & 0 \\
0 & 0 & 0 & \veps
\end{array}\right)
h_1\left(\br+\bL_2\right).
\end{eqnarray}
So,
\begin{eqnarray}\label{eqn:Apx P on h1}
-i\mu_y w_{\bR,1}(-\br)
&=&-
\frac{1}{\sqrt{N_{uc}}}\sum_{m,\bk} e^{i(\bk+\bq_1)\cdot\bR} \Phi_{m,\bk}(\br)\text{UP}\left[\int d^2\br' \Phi^\dagger_{m,\bk}(\br') \left(\begin{array}{cccc}
\veps^* & 0 & 0 & 0\\
0 & \veps^* & 0 & 0 \\
0 & 0 & \veps & 0 \\
0 & 0 & 0 & \veps
\end{array}\right)
h_2\left(\br'+\bL_2\right)\right]\\
\label{eqn:Apx P on h2}
-i\mu_y w_{\bR,2}(-\br)
&=&
\frac{1}{\sqrt{N_{uc}}}\sum_{m,\bk} e^{i(\bk+\bq_1)\cdot\bR} \Phi_{m,\bk}(\br)\text{UP}\left[\int d^2\br' \Phi^\dagger_{m,\bk}(\br') \left(\begin{array}{cccc}
\veps^* & 0 & 0 & 0\\
0 & \veps^* & 0 & 0 \\
0 & 0 & \veps & 0 \\
0 & 0 & 0 & \veps
\end{array}\right)
h_1\left(\br'+\bL_2\right)\right]
\end{eqnarray}
Shifting the overlap integral and using the definition (\ref{eqn:Psi}) we have
\begin{eqnarray}
&&\int d^2\br' \Phi^\dagger_{m,\bk}(\br') \left(\begin{array}{cccc}
\veps^* & 0 & 0 & 0\\
0 & \veps^* & 0 & 0 \\
0 & 0 & \veps & 0 \\
0 & 0 & 0 & \veps
\end{array}\right)
h_j\left(\br'+\bL_2\right)=
\int d^2\br' \Phi^\dagger_{m,\bk}(\br'-\bL_2) \left(\begin{array}{cccc}
\veps^* & 0 & 0 & 0\\
0 & \veps^* & 0 & 0 \\
0 & 0 & \veps & 0 \\
0 & 0 & 0 & \veps
\end{array}\right)
h_j\left(\br'\right)\\
&=&e^{i\bk\cdot\bL_2}\int d^2\br' \Phi^\dagger_{m,\bk}(\br') \left(\begin{array}{cccc}
1 & 0 & 0 & 0\\
0 & 1 & 0 & 0 \\
0 & 0 & e^{i\bq_1\bL_2} & 0 \\
0 & 0 & 0 & e^{i\bq_1\bL_2}
\end{array}\right)\left(\begin{array}{cccc}
\veps^* & 0 & 0 & 0\\
0 & \veps^* & 0 & 0 \\
0 & 0 & \veps & 0 \\
0 & 0 & 0 & \veps
\end{array}\right)
h_j\left(\br'\right)
=\veps^* e^{i\bk\cdot\bL_2} \int d^2\br' \Phi^\dagger_{m,\bk}(\br') h_j\left(\br'\right)
\end{eqnarray}
Substituting into (\ref{eqn:Apx P on h1}) and (\ref{eqn:Apx P on h2}) we finally have
\begin{eqnarray}
-i\mu_y w_{\bR,1}(-\br)
&=&-\veps^* e^{i\bq_1\cdot\bR}w_{-\bR-\bL_2,2}(\br)\\
-i\mu_y w_{\bR,2}(-\br)
&=&\veps^* e^{i\bq_1\cdot\bR}w_{-\bR-\bL_2,1}(\br)
\end{eqnarray}
which match the result stated in the main text (\ref{eqn:P on w1}) and (\ref{eqn:P on w2}).

\subsection{Action of the symmetry operation $\mathcal{C}C_2T$}
\begin{eqnarray}
\sigma_z\sigma_x w^*_{\bR,j}(-\br)&=&
\frac{1}{\sqrt{N_{uc}}}\sum_{m,\bk} e^{i\bk\cdot\bR} \sigma_z\sigma_x\Phi^*_{m,\bk}(-\br)\text{UP}\left[\int d^2\br' \Phi^\dagger_{m,\bk}(\br')h_{j}(\br')\right]^*\\
&=&
\frac{1}{\sqrt{N_{uc}}}\sum_{m,\bk} e^{i\bk\cdot\bR} i\sigma_y\Phi^*_{m,\bk}(-\br)\text{UP}\left[\int d^2\br' \Phi^T_{m,\bk}(-\br')(-i\sigma_y)i\sigma_yh^*_{j}(-\br')\right]\\
&=&
\frac{1}{\sqrt{N_{uc}}}\sum_{m,\bk} e^{i\bk\cdot\bR} \Phi_{m,\bk}(\br)\text{UP}\left[\int d^2\br' \Phi^\dagger_{m,\bk}(\br')i\sigma_yh^*_{j}(-\br')\right]
\end{eqnarray}

\begin{eqnarray}
i\sigma_y h^*_1(-\br)&=& \left(\begin{array}{c} 0 \\ i \\  \veps \\0
\end{array}\right)\delta\left(\br+\br_{AB}\right)=
\left(\begin{array}{cccc}
-i & 0 & 0 & 0 \\
0 & -i & 0 & 0 \\
0 & 0 & -i\veps^* & 0 \\
0 & 0 & 0 & -i\veps^*
\end{array}\right)
h_2\left(\br+\bL_2\right)\\
i\sigma_y h^*_2(-\br)&=&
\left(\begin{array}{c} -1 \\ 0 \\ 0 \\ -i\veps
\end{array}\right)\delta\left(\br+\br_{BA}\right)=
\left(\begin{array}{cccc}
i & 0 & 0 & 0 \\
0 & i & 0 & 0 \\
0 & 0 & i\veps^* & 0 \\
0 & 0 & 0 & i\veps^*
\end{array}\right)
h_1\left(\br+\bL_2\right)
\end{eqnarray}

\begin{eqnarray}
&&\sigma_z\sigma_x w^*_{\bR,1(2)}(-\br)=
\frac{\mp i}{\sqrt{N_{uc}}}\sum_{m,\bk} e^{i\bk\cdot\bR} \Phi_{m,\bk}(\br)\text{UP}\left[\int d^2\br' \Phi^\dagger_{m,\bk}(\br')
\left(\begin{array}{cccc}
1 & 0 & 0 & 0 \\
0 & 1 & 0 & 0 \\
0 & 0 & \veps^* & 0 \\
0 & 0 & 0 & \veps^*
\end{array}\right)
h^*_{2(1)}(\br'+\bL_2)\right]\\
&=&\frac{\mp i}{\sqrt{N_{uc}}}\sum_{m,\bk} e^{i\bk\cdot(\bR+\bL_2)} \Phi_{m,\bk}(\br)\text{UP}\left[\int d^2\br' \Phi^\dagger_{m,\bk}(\br')
\left(\begin{array}{cccc}
1 & 0 & 0 & 0 \\
0 & 1 & 0 & 0 \\
0 & 0 & e^{i\bq_1\cdot\bL_2} & 0 \\
0 & 0 & 0 & e^{i\bq\cdot\bL_2}
\end{array}\right)
\left(\begin{array}{cccc}
1 & 0 & 0 & 0 \\
0 & 1 & 0 & 0 \\
0 & 0 & \veps^* & 0 \\
0 & 0 & 0 & \veps^*
\end{array}\right)
h^*_{2(1)}(\br')\right]\\
\\
&=&\mp iw_{-\bR-\bL_2,2(1)}(\br).
\end{eqnarray}
This matches the result in Eqns.~\ref{eqn:CC2T on w1} and ~\ref{eqn:CC2T on w2}.

\section{Relation between different $U(4)$ rotations}
\label{apx:U4s}

In order to make explicit the relation between different $U(4)$ symmetries mentioned in the main text, we start by recalling the definition of the fermion operators in different bases 
\begin{eqnarray}
\chi_\sigma(\br)&=&
\left(\begin{array}{c} \psi_\sigma(\br) \\ \phi_\sigma(\br)\end{array}\right)
=\sum_{\bk}\sum_{n=n\pm}
\left(\begin{array}{c} \Psi_{n,\bk}(\br) d_{\sigma,\bK,n,\bk} \\ \Psi^*_{n,\bk}(\br) d_{\sigma,\bK',n,-\bk-\bq_1}
\end{array}\right)=\sum_{\bk}\sum_{\lambda=\pm}
\left(\begin{array}{c} \Phi_{\lambda,\bk}(\br) e_{\sigma,\bK,\lambda,\bk} \\ e^{i\alpha^{\mathcal{C}}_\bk}\Phi^*_{\lambda,\bk}(\br) e_{\sigma,\bK',-\lambda,-\bk-\bq_1}
\end{array}\right)\nonumber\\
&=&
\sum_{\bR}\sum_{j=1,2}
\left(\begin{array}{c} \tilde{w}_{\bR,j}(\br) f_{\sigma,\bK,j,\bR} \\ \tilde{w}^*_{\bR,j}(\br) f_{\sigma,\bK',j,\bR}
\end{array}\right).
\end{eqnarray}
Relation between the basis functions are
\begin{eqnarray}
\Phi_{\pm,\bk}(\br)&=&\frac{1}{\sqrt{2}}\left(1_2\sigma_z \pm 1_4\right)\Psi_{n+,\bk}(\br)=\frac{1}{\sqrt{2}}\left(e^{i\alpha^\mathcal{C}_\bk}\Psi_{n-,\bk}(\br) \pm \Psi_{n+,\bk}(\br)\right)\\
\tilde{w}_{\bR,j}(\br)&=&e^{i\frac{\pi}{12}}e^{-\frac{i}{2}\bq_1\cdot\bR} w_{\bR,j}(\br)=
\frac{1}{\sqrt{N_{uc}}}e^{i\frac{\pi}{12}}\sum_{\bk}\sum_{m=\pm} e^{-i\left(\bk+\frac{\bq_1}{2}\right)\cdot\bR}
\Phi_{m,\bk}(\br)\mathcal{U}_{mj}(\bk)
\end{eqnarray}

Let's start with the first equality, and read off the coefficients using the orthogonality of $\Psi's$.
\begin{eqnarray}
d_{\sigma,\bK,n+,\bk}&=&\frac{1}{\sqrt{2}}\left(e_{\sigma,\bK,+,\bk}-e_{\sigma,\bK,-,\bk}\right)\\
d_{\sigma,\bK,n-,\bk}&=&\frac{e^{i\alpha^\mathcal{C}_\bk}}{\sqrt{2}}\left(e_{\sigma,\bK,+,\bk}+e_{\sigma,\bK,-,\bk}\right)\\
d_{\sigma,\bK',n+,-\bk-\bq_1}&=&\frac{e^{i\alpha^{\mathcal{C}}_\bk}}{\sqrt{2}}\left(e_{\sigma,\bK',-,-\bk-\bq_1}-e_{\sigma,\bK',+,-\bk-\bq_1}\right)\\
d_{\sigma,\bK',n-,-\bk-\bq_1}&=&\frac{1}{\sqrt{2}}\left(e_{\sigma,\bK',-,-\bk-\bq_1}+e_{\sigma,\bK',+,-\bk-\bq_1}\right)
\end{eqnarray}
where we used
\begin{eqnarray}
\Psi^*_{n+,\bk}(\br) d_{\sigma,\bK',n+,-\bk-\bq_1}+\Psi^*_{n-,\bk}(\br) d_{\sigma,\bK',n-,-\bk-\bq_1}=
e^{i\alpha^{\mathcal{C}}_\bk}\Phi^*_{+,\bk}(\br) e_{\sigma,\bK',-,-\bk-\bq_1}+
e^{i\alpha^{\mathcal{C}}_\bk}\Phi^*_{-,\bk}(\br) e_{\sigma,\bK',+,-\bk-\bq_1}.
\end{eqnarray}

\begin{eqnarray}
\left(\begin{array}{c}
d_{\sigma,\bK,n+,\bk}\\
d_{\sigma,\bK,n-,\bk}\\
d_{\sigma,\bK',n+,\bk}\\
d_{\sigma,\bK',n-,\bk}
\end{array}\right)=\frac{1}{\sqrt{2}}
\left(\begin{array}{cccc}
1 & -1 & 0 & 0\\
e^{i\alpha^\mathcal{C}_\bk} & e^{i\alpha^\mathcal{C}_\bk} & 0 & 0 \\
0 & 0 & e^{i\alpha^\mathcal{C}_{-\bk-\bq_1}} & -e^{i\alpha^\mathcal{C}_{-\bk-\bq_1}}\\
0 & 0 & 1 & 1
\end{array}\right)
\left(\begin{array}{c}
e_{\sigma,\bK,+,\bk}\\
e_{\sigma,\bK,-,\bk}\\
e_{\sigma,\bK',-,\bk}\\
e_{\sigma,\bK',+,\bk}
\end{array}\right)
\end{eqnarray}
Now, if $\bk=\fvec{\Gamma}$ then $e^{i\alpha^\mathcal{C}_\bk}=e^{i\alpha^\mathcal{C}_{-\bk-\bq_1}}=i$.
If $\bk \neq \fvec{\Gamma}$ then $e^{i\alpha^\mathcal{C}_\bk}=e^{i\alpha^\mathcal{C}_{-\bk-\bq_1}}=-i$

By explicit calculation we find that 
\begin{eqnarray}
\left[G^{nc-flat}_{U(4)},\frac{1}{\sqrt{2}}
\left(\begin{array}{cccc}
1 & -1 & 0 & 0\\
e^{i\alpha^\mathcal{C}_\bk} & e^{i\alpha^\mathcal{C}_\bk} & 0 & 0 \\
0 & 0 & e^{i\alpha^\mathcal{C}_{-\bk-\bq_1}} & -e^{i\alpha^\mathcal{C}_{-\bk-\bq_1}}\\
0 & 0 & 1 & 1
\end{array}\right)
\right]=0
\end{eqnarray}
where $G^{nc-flat}_{U(4)}$ is any of the non-chiral flat generators $1_4$, $\tau_z 1_2$, $\tau_y\tilde{\sigma}_y$, or $\tau_x\tilde{\sigma}_y$.
Writing out these matrices explicitly we can see that if we rotate the Chern +1 basis using 
$\alpha\cdot\left(1_2,\tau_z,-\tau_x,\tau_y\right)$ then we must simultaneously rotate Chern -1 using 
$\alpha\cdot\left(1_2,\tau_z,\tau_x,-\tau_y\right)$. The spin SU(2) follows trivially from the tensor product.

The second equality implies
\begin{eqnarray}
f_{\sigma,\bK,j,\bR}&=&
\sum_{\bk}\sum_{\lambda=\pm} \int d^2\br \tilde{w}^\dagger_{\bR,j}(\br)\Phi_{\lambda,\bk}(\br) e_{\sigma,\bK,\lambda,\bk}\\
f_{\sigma,\bK',j,\bR}&=&\sum_{\bk}\sum_{\lambda=\pm}e^{i\alpha^{\mathcal{C}}_\bk} \int d^2\br \tilde{w}^T_{\bR,j}(\br)\Phi^*_{\lambda,\bk}(\br) e_{\sigma,\bK',-\lambda,-\bk-\bq_1}
\end{eqnarray}
Therefore,
\begin{eqnarray}
f_{\sigma,\bK,j,\bR}&=&
\frac{1}{\sqrt{N_{uc}}}\sum_{\bk}\sum_{\lambda=\pm} 
e^{-i\frac{\pi}{12}} e^{i\left(\bk+\frac{\bq_1}{2}\right)\cdot\bR}
\mathcal{U}^*_{\lambda j}(\bk)
 e_{\sigma,\bK,\lambda,\bk}\\
f_{\sigma,\bK',j,\bR} &=&\frac{1}{\sqrt{N_{uc}}}\sum_{\bk}\sum_{\lambda=\pm}e^{i\alpha^{\mathcal{C}}_\bk} 
e^{i\frac{\pi}{12}} e^{-i\left(\bk+\frac{\bq_1}{2}\right)\cdot\bR}
\mathcal{U}_{\lambda j}(\bk)
 e_{\sigma,\bK',-\lambda,-\bk-\bq_1}\\
&=&\frac{1}{\sqrt{N_{uc}}}\sum_{\bk}\sum_{\lambda=\pm}e^{i\alpha^{\mathcal{C}}_{-\bk-\bq_1}}
e^{i\frac{\pi}{12}} e^{i\left(\bk+\frac{\bq_1}{2}\right)\cdot\bR}
\mathcal{U}_{-\lambda, j}(-\bk-\bq_1)
 e_{\sigma,\bK',\lambda,\bk}
\end{eqnarray}
Now,
\begin{eqnarray}
e^{i\alpha^{\mathcal{C}}_{-\bk-\bq_1}}\mathcal{U}_{-\lambda, j}(-\bk-\bq_1)&=&e^{i\alpha^{\mathcal{C}}_{-\bk-\bq_1}}\text{UP}
\left[\int d^2\br \Phi^\dagger_{-\lambda,-\bk-\bq_1}(\br)h_j(\br)\right]
\end{eqnarray}
But, 
\begin{eqnarray}
e^{i\alpha^{\mathcal{C}}_{-\bk-\bq_1}}\Phi^\dagger_{-\lambda,-\bk-\bq_1}(\br)=-\Phi^T_{\lambda,\bk}(\br)\mu_y\sigma_y
\end{eqnarray}
So,
\begin{eqnarray}
e^{i\alpha^{\mathcal{C}}_{-\bk-\bq_1}}\mathcal{U}_{-\lambda, j}(-\bk-\bq_1)&=&-\text{UP}
\left[\int d^2\br \Phi^\dagger_{\lambda,\bk}(\br)\mu_y\sigma_y h^*_j(\br)\right]^*
\end{eqnarray}
and for our choice of the trial states
\begin{eqnarray}
\mu_y\sigma_y h_j^*(\br)&=&e^{i\frac{\pi}{6}}h_j(\br).
\end{eqnarray}
Therefore,
\begin{eqnarray}
e^{i\alpha^{\mathcal{C}}_{-\bk-\bq_1}}\mathcal{U}_{-\lambda, j}(-\bk-\bq_1)&=&-e^{-i\frac{\pi}{6}}\text{UP}
\left[\int d^2\br \Phi^\dagger_{\lambda,\bk}(\br) h_j(\br)\right]^*=-e^{-i\frac{\pi}{6}}\mathcal{U}^*_{\lambda,j}(\bk)
\end{eqnarray}
Therefore we get,
\begin{eqnarray}
f_{\sigma,\bK,j,\bR}&=&
\frac{1}{\sqrt{N_{uc}}}\sum_{\bk}\sum_{\lambda=\pm}
e^{-i\frac{\pi}{12}} e^{i\left(\bk+\frac{\bq_1}{2}\right)\cdot\bR}
\mathcal{U}^*_{\lambda j}(\bk)
 e_{\sigma,\bK,\lambda,\bk}\\
f_{\sigma,\bK',j,\bR} 
&=&-\frac{1}{\sqrt{N_{uc}}}\sum_{\bk}\sum_{\lambda=\pm}
e^{i\frac{\pi}{12}} e^{i\left(\bk+\frac{\bq_1}{2}\right)\cdot\bR}
e^{-i\frac{\pi}{6}}\mathcal{U}^*_{\lambda,j}(\bk)
 e_{\sigma,\bK',\lambda,\bk}
\end{eqnarray}
Or, expressed in a matrix form
\begin{eqnarray}
\left(\begin{array}{c}
f_{\sigma,\bK,j,\bR}\\ f_{\sigma,\bK',j,\bR}
\end{array}
\right)&=&\frac{1}{\sqrt{N_{uc}}}\sum_{\bk}\sum_{\lambda=\pm}
e^{-i\frac{\pi}{12}} e^{i\left(\bk+\frac{\bq_1}{2}\right)\cdot\bR}
\mathcal{U}^*_{\lambda j}(\bk)
\left(\begin{array}{cc}
1 & 0 \\ 0 & -1
\end{array}\right)
\left(\begin{array}{c}
e_{\sigma,\bK,\lambda,\bk}\\ e_{\sigma,\bK',\lambda,\bk}
\end{array}\right)
\end{eqnarray}

So, if we perform a valley $U(2)$ rotation $e^{i\alpha\cdot(1,\tau_z,\tau_x,\tau_y)}$ on $f$'s, then we are performing  
$e^{i\alpha\cdot(1,\tau_z,-\tau_x,-\tau_y)}$ rotation on $e$'s {\it independent} of $\lambda$.
Clearly, the two Chern numbers are rotated by the same angle. This is unlike in the previous example of the non-chiral flat limit where the Chern +1 and Chern -1 were rotated by angle vectors which are related by mirror reflection about the $z$-axis.

\end{widetext}

\newpage

\end{document}